# One-step Preparation of ZnO Electron Transport Layers Functionalized with Benzoic Acid Derivatives

Hao Liu,[a] Chao Wang,[a] Wen Liang Tan,[a] Lars Thomsen,[b] Anthony S. R. Chesman,[c] Yvonne Hora,[d] Martyn Jevric,[e] Jonas M. Bjuggren,[e] Mats R. Andersson,[e] Yahui Tang,[a] Linjing Tang,[a] Doan Vu,[a] and Christopher R. McNeill[a]*

[a]Department of Materials Science and Engineering, Monash University, Wellington Road, Clayton, Victoria, 3800, Australia
[b]Australian Synchrotron, ANSTO, 800 Blackburn Road, Clayton, Victoria, 3168, Australia.
[c]CSIRO Manufacturing, Ian Wark Laboratories, Clayton, Victoria 3168, Australia
[d]Monash X-ray Platform, Monash University, Wellington Road, Clayton, Victoria, 3800, Australia
[e]Flinders Institute for NanoScale Science and Technology, Flinders University, Adelaide, South Australia, 5042, Australia
*Corresponding author. E-mail: christopher.mcneill@monash.edu

**Abstract:**
We present a "one-step" approach to modify ZnO electron transport layers (ETLs) used in organic solar cells. This approach involves adding benzoic acid (BZA) derivatives directly to the ZnO precursor solution, which are then present at the surface of the resulting ZnO film. We demonstrate this approach for three different BZA derivatives, namely benzoic acid, chlorobenzoic acid, and 4-hydrazinobenzoic acid. For all molecules, improved device performance and stability is demonstrated in solar cells using an active layer blend of PTQ10 (donor) and ITIC-Br (non-fullerene acceptor) compared to such cells prepared using untreated ZnO. Furthermore, similar or improved device performance and stability is demonstrated compared to conventional PEIE treatment of ZnO. The presence of the BZA derivatives at the surface after processing is established using X-ray photoelectron spectroscopy and near-edge X-ray absorption fine-structure spectroscopy. From atomic force microscopy analysis and X-ray diffraction studies, the addition of BZA derivatives appears to restrict ZnO grain growth; however, this does not negatively impact device performance. ZnO layers treated with BZA derivatives also exhibit higher water contact angle and lower work function compared to untreated ZnO. This approach enables simplification of device manufacture while still allowing optimization of the surface properties of metal oxide ETLs.

**Graphical Abstract:**

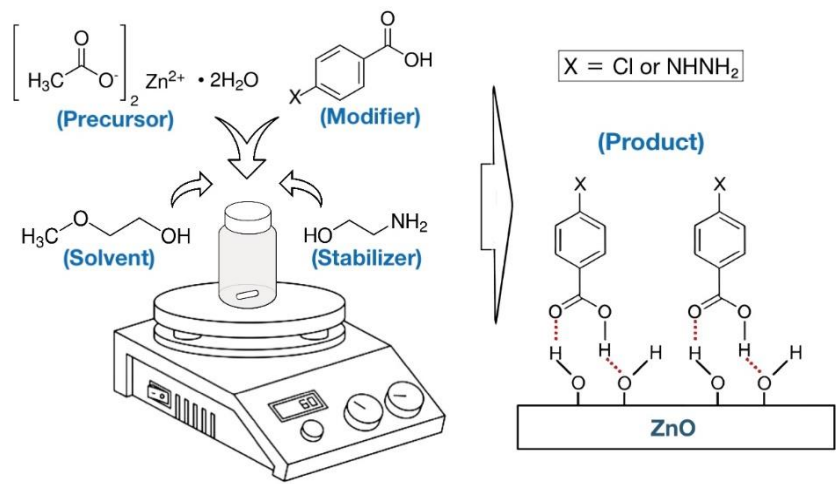

## 1. Introduction

The realm of photovoltaics continues to receive growing attention due to rising demand for clean energy. Next generation photovoltaic technologies, such as organic solar cells (OSCs), are particularly interesting due to their unique advantages such as comparably low-cost and ease of large-area fabrication, light weight, mechanical flexibility, and potential for semi-transparent application [1-3]. In recent years, the record power conversion efficiency (PCE) of single-junction and tandem organic solar cells (OSCs) has surpassed 19% and 20%, respectively, bringing the technology closer to commercialization [4, 5]. This rapid progress can be largely attributed to the development of non-fullerene acceptors (NFAs), as well as device engineering including surface or interface engineering [6, 7].

Efficient solar cell operation relies on high-quality electrode interfaces characterized by Ohmic contact and high charge selectivity [8, 9]. However, owing to the nanostructured morphology of the active layer, interface optimization can be difficult to achieve in bulk heterojunction (BHJ) OSCs [9]. To address this matter, a common approach is to employ additional interlayers between the BHJ and electrodes, namely the electron transport layer (ETL) and hole transport layer (HTL). These charge transport layers help to pin the Fermi levels of the electrodes to the quasi-Fermi levels of the charges in the BHJ under illumination, thereby alleviating the surface energy barrier in between [9]. For instance, an ideal ETL needs to have a relatively low work function to align with the quasi-Fermi levels of electrons ($E_{F,e^-}$) in the BHJ. For molecular ETLs, the LUMO of the ETL should align with $E_{F,e^-}$, while the highest occupied molecular orbital (HOMO) of the ETL needs to be distinctly lower than that of the quasi-Fermi levels of holes ($E_{F,h+}$) of the BHJ to ensure that the cathode only collects electrons while blocking holes (i.e. electron selectivity), thereby avoiding unfavorable charge flow and accumulation [8, 9].

As the main purpose of the charge transport layers in OSCs is to promote the extraction of charges while minimizing charge recombination, choice of interlayer material and interfacial modification method plays a significant role in determining device performance and stability [2, 8]. Choice of ETL and HTL is more complicated but crucial for non-fullerene-based systems than for fullerene-based counterparts, because the isotropic geometry of fullerene derivatives promotes effective π-π interactions for charge transfer [7]. Moreover, the choice of interlayer can also affect the stability of OSCs, with the charge transport layer acting as a protection layer between the active layer and the electrodes to impede the diffusion of material from the metal electrodes into devices, as well as preventing the permeation of external moisture and oxygen into the BHJ film [8-10].

There have been numerous types of materials used as ETLs, including small molecules or oligomers such as fullerene or non-fullerene-based derivatives; (neutral) polymers like PEIE (polyethylenimine ethoxylated) or PEI (polyethylenimine); polyelectrolytes; low work function metals like Ca, Ba, Mg and Al, and their salts or complexes; carbon-based materials including graphene oxides; transition or semiconducting metal oxides including ZnO, $TiO_2$ and $In_2O_3$; and organic-inorganic hybrids or composites such as ZnO/PEIE [8, 9]. Among all these types, metal oxides are among the most frequently used ETL materials because they are able to strike a balance between solution processability, ambient stability and favorable electronic properties [8]. ZnO in particular, being a solution-processable inorganic n-type semiconductor, is a common choice as ETL material owing to its low cost, ease of preparation, non-toxicity, high stability, and good optical and electronic properties [2, 11]. ZnO possesses a low work function of about 4.30 eV, which provides a suitable energy level for reducing the work function of ITO or metal electrodes, thereby better matching with the LUMO levels of potential acceptors [8, 9]. ZnO films are typically prepared via a low-temperature sol-gel reaction [12], as is employed in this article. Other ZnO preparation routes include electrochemical deposition, atomic layer deposition, nanoparticle approaches, and hydrothermal growth [13-16]. Key ETL properties of ZnO, such as electron mobility and surface energy, are dependent on properties such as morphology, composition, crystalline structure and defects, and layer thickness, which can be adjusted by varying processing conditions [17-22]. As such, OSC performance can be enhanced by modifying the ZnO used as the ETL, which emphasizes the importance of managing the properties of the ETL material to attain an optimal combination of mobility, interface conditions and other properties in order to create high-performing OSCs [8].

Previous studies have shown that ETLs based on ZnO can be improved by functionalizing with an additional organic interlayer. A standard ETL modification approach is to coat ZnO with a very thin layer of PEIE to form a composite ETL [2, 8, 9]. This approach enhances charge extraction at the interface between the ETL through the effect of an additional dipole layer provided by the PEIE coating [23, 24]. Another popular approach is to employ a functionalized molecule, such as benzoic acid (BZA), which is often coated on the ZnO as a self-assembled monolayer (SAM) [3, 25-28]. This approach similarly enhances selective electron collection at the interface by reducing the apparent work function of the n-type metal oxide ETL via an additional dipole interlayer [25-27]. The preparation of such modified ZnO ETLs required at least two coating steps during the preparation process, for example, deposition of the ZnO layer followed by deposition of PEIE or SAM. This two-step approach adds to the complexity of the device manufacturing process. An alternative approach which is demonstrated here, is adding the interface modifying molecule directly into the ZnO precursor solution to synthesize a hybrid ETL material produced by a one-step method.

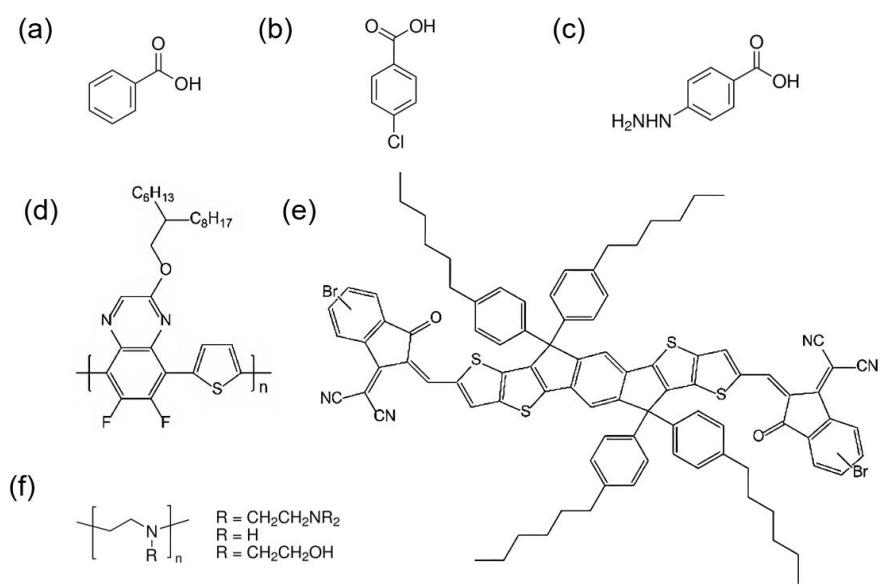

Figure 1. Chemical structures of materials used in this study: (a) Benzoic acid (BZA), (b) 4-Chlorobenzoic acid (CBA), (c) 4-Hydrazinobenzoic acid (HBA), (d) PTQ10 (donor), (e) ITIC-Br (acceptor), (f) polyethylenimine ethoxylated (PEIE).

## 2. Results and Discussion

### 2.1. ETL Preparation

The ZnO ETLs used in this study were prepared using a sol-gel route, where a 0.1 M solution of zinc acetate dihydrate in 2-methoxyethanol was prepared with a small amount of ethanolamine added as stabilizer (see experimental section for further details). For BZA-modified ZnO ETLs, 0.01 M of BZA derivatives, namely benzoic acid (BZA), chlorobenzoic acid (CBA), and hydrazinobenzoic acid (HBA) (see Figure 1 for chemical structures) was added to the precursor solution. The solution was continuously stirred on a hotplate at 60 °C for 20 hours for hydrolysis followed by filtration prior to layer deposition. ZnO layers were prepared by spin coating followed by annealing under ambient conditions at 200 °C for 30 minutes.

### 2.2. Photovoltaic Performance

Devices used in this study have an active layer composed of the low-cost donor polymer (PTQ10) and the novel non-fullerene acceptor (ITIC-Br) [29, 30], see Figure 1 for chemical structures. Devices had an inverted structure of ITO/ZnO/PTQ10:ITIC-Br/MoO$_x$/Ag with the ZnO layer modified as described above. For comparison, devices with a thin PEIE layer coated on top of the ZnO were also prepared.

The JV curves of solar cells fabricated with different ETLs under AM 1.5G simulated illumination with an intensity of 100 mW/cm$^2$ are presented in Figure 2, with the corresponding photovoltaic parameters summarized in Table 1. Compared to the reference ZnO-only devices (i.e. unmodified ZnO), modifying ZnO with BZA derivative improves all photovoltaic parameters. The reference ZnO-only devices show an average power conversion efficiency (PCE) of 9.4%, open circuit voltage ($V_{OC}$) of 0.88 V, short circuit current ($J_{SC}$) of 19.8 mA/cm$^2$, and a fill factor (FF) of 0.53. Coating ZnO with a thin layer of PEIE improves the PCE to 10.6%, consistent with prior work [31]. The enhancement in device performance is mainly due to an increase in FF from 0.53 to 0.59, and a slight increase in $V_{OC}$ from 0.88 V to 0.92 V. On the other hand, $J_{SC}$ showed a slight decrease with respect to the ZnO reference device. For devices modified with the various BZA derivatives, a similar improvement in FF up to 0.6 and improvement in $V_{OC}$ up to 0.92 V is observed. Interestingly, $J_{SC}$ is also improved in the BZA-treated devices, increasing from 19.8 mA/cm$^2$ for the ZnO reference devices to 20.1 mA/cm$^2$ for the BZA-treated devices, 20.4 mA/cm$^2$ for the CBA treated device, and 20.5 mA/cm$^2$ for the HBA-treated devices. Among all the BZA derivatives employed here, the HBA modified ZnO ETL achieves the highest PCE, which shows an improvement in PCE from 9.4% for the untreated reference ZnO ETL to 11.3%. The BZA and CBA modified devices exhibit lower overall PCEs, but still higher than that of the PEIE-modified device.

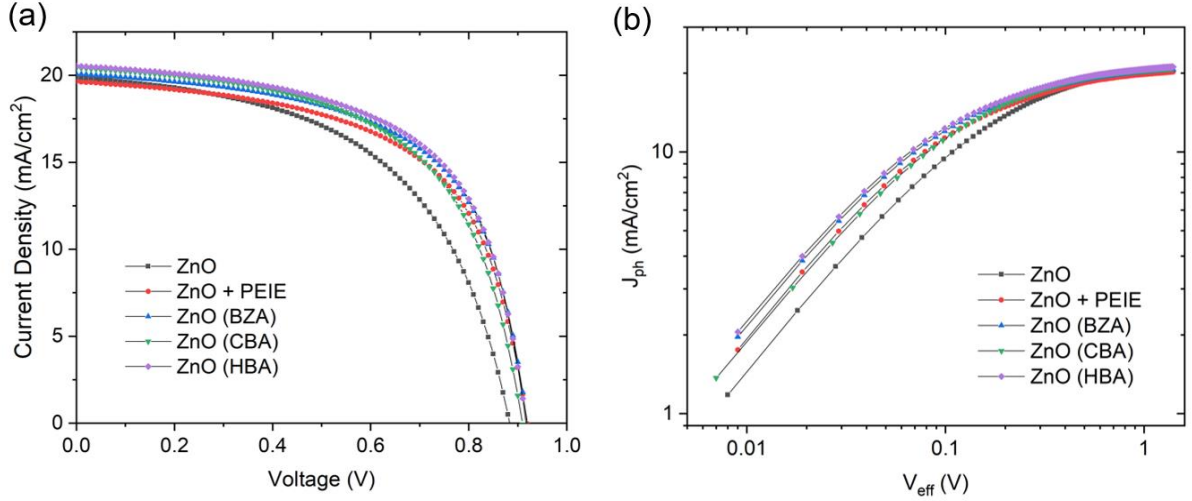

Figure 2. (a) Light JV curves of PTQ10:ITIC-Br OSCs based on various ETLs; (b) Photocurrent density ($J_{ph}$) as a function of the effective voltage ($V_{eff}$) for the OSCs with various ETLs.

Table 1. Average photovoltaic parameters of PTQ10:ITIC-Br based solar cells with different interface modification strategies. The mean values and standard deviations for each ETL configuration were derived from 5 sets of samples.

| ETL | $V_{OC}$ (V) | $J_{SC}$ (mA/cm$^2$) | FF | PCE (%) |
|---|---|---|---|---|
| ZnO (Reference) | 0.88 (± 0.01) | 19.8 (± 0.4) | 0.53 (± 0.01) | 9.4 (± 0.4) |
| ZnO + PEIE | 0.92 (± 0.01) | 19.6 (± 0.4) | 0.59 (± 0.01) | 10.6 (± 0.4) |
| ZnO + BZA | 0.92 (± 0.01) | 20.1 (± 0.5) | 0.60 (± 0.01) | 11.1 (± 0.5) |
| ZnO + CBA | 0.91 (± 0.01) | 20.4 (± 0.8) | 0.58 (± 0.03) | 10.8 (± 0.5) |
| ZnO + HBA | 0.92 (± 0.01) | 20.5 (± 0.8) | 0.60 (± 0.01) | 11.3 (± 0.3) |

The improvement in electron extraction efficiency at the ETL interfaces is also verified by comparing the dependence of photocurrent density ($J_{ph}$) with effective voltage ($V_{eff}$) for different ZnO configurations, as shown in Figure 2 (b). $J_{ph}$ and $V_{eff}$ are defined by the following equations:

$$J_{ph} = J_L - J_D \tag{1}$$

where $J_L$ and $J_D$ are the light current density and dark current density, respectively;

$$V_{eff} = V_0 - V \tag{2}$$

where $V_0$ is the compensation voltage at which $J_{ph}$ is zero, and V is the applied voltage [32]. According to Figure 2 (b), it can be seen that $J_{ph}$ increases linearly with $V_{eff}$ at values less than 0.1 V, while at values above 1 V $J_{ph}$ tends to saturate. The saturated value of $J_{ph}$ is the same for all devices, limited by the absorbed photo flux [32, 33]. In contrast there are differences in the $J_{ph}$ values in the low $V_{eff}$ range, which are associated with different electron extraction efficiencies. In general, a higher $J_{ph}$ represents a higher charge extraction efficiency, thereby leading to a higher FF of the device [33]. As the variation in photocurrent extraction efficiency at low $V_{eff}$ corresponds well to the observed FF variations, it can be reasonably inferred that improved interfacial electron extraction is one of the key factors for the increase in FF, and hence PCE.

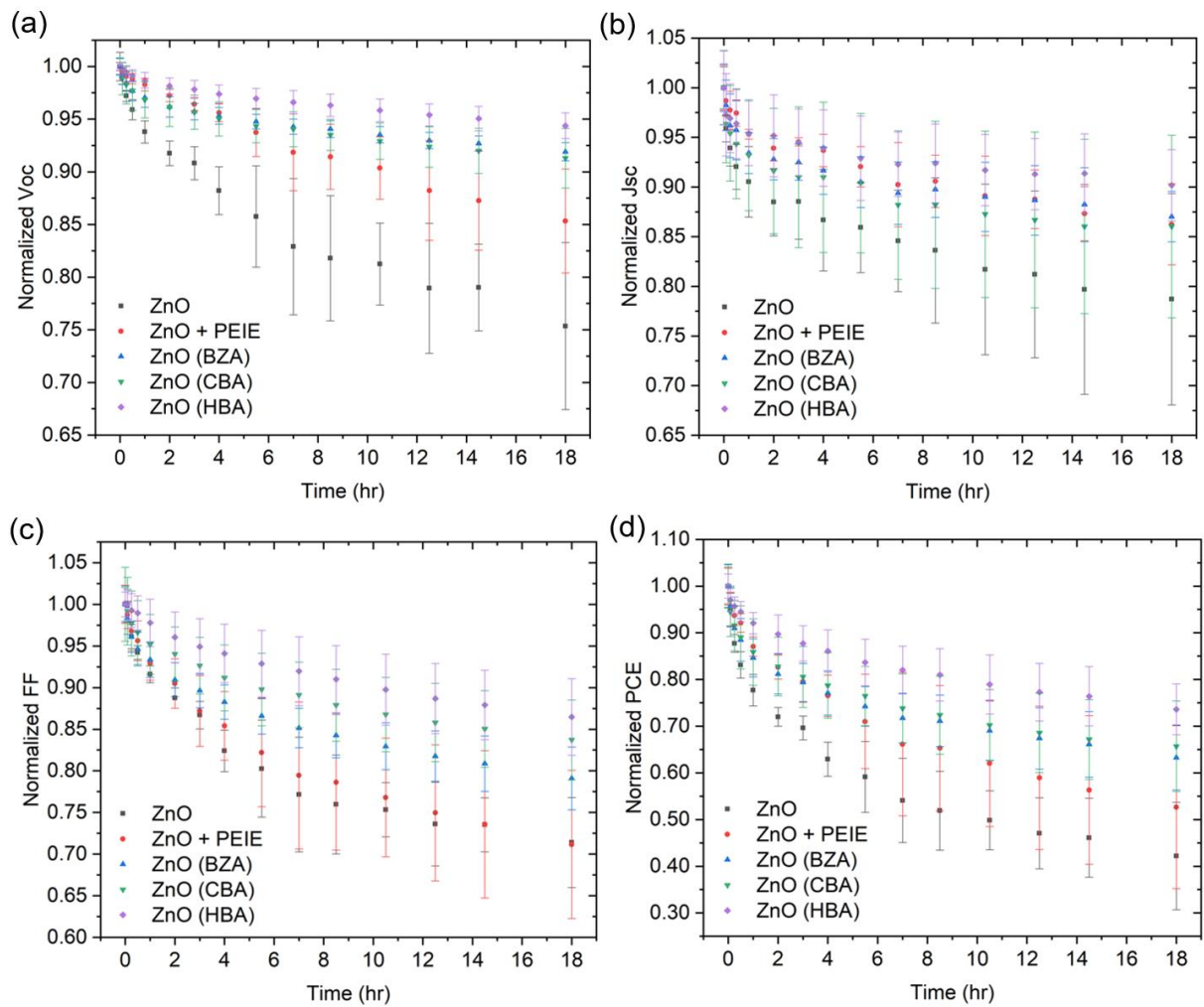

Figure 3. Normalized photovoltaic parameters of PTQ10:ITIC-Br based OSCs as a function of light soaking time, prepared with different ETLs: (a) $V_{OC}$; (b), $J_{SC}$; (c), FF; (d) PCE. The mean and standard deviation for each ETL configuration were derived from 5 devices.

**2.3. Operational Stability**

Degradation studies under constant illumination show that the operational stability of devices with modified ZnO are improved. Figure 3 presents normalized plots of the four key photovoltaic parameters (i.e. $V_{OC}$, $J_{SC}$, FF and PCE) as a function of time under continuous light soaking. The stability tests were conducted according to the ISOS-L-1 laboratory weathering testing protocol, where devices were placed under the solar simulator at 100 mW/cm$^2$ intensity and tested after set periods of time [34]. To aid discussion, the "$T_{80}$" metric is adopted to quantify operational stability, which corresponds to time taken for the performance to degrade to 80% of the initial performance [34]. From Figure 3 it can be seen that different ETLs result in different device lifetimes. The reference ZnO device has the worst operational stability with a $T_{80}$ (PCE) of less than 1 hour. The BZA, CBA and PEIE modified devices have similar $T_{80}$ values of around 3 hours, while the HBA-modified device has a $T_{80}$ of nearly 9 hours. Going beyond $T_{80}$, the BZA, CBA and PEIE modified device curves diverge with the CBA device retaining 65% of its initial efficiency after 18 hours compared to 62% for the BZA device, 52% for the PEIE device with the reference ZnO device only retaining 42% of its initial efficiency. The HBA device however shows the best stability overall retaining around 74% of its initial efficiency after 18 hours. These results show that modification of the ZnO surface can improve device stability with the BZA derivatives providing superior device stability to PEIE.

**2.4 Characterization of Modified ZnO Layers**

*2.4.1. X-ray Photoelectron Spectroscopy (XPS) Analysis*

From the above results, it is clear that the addition of BZA derivatives to the ZnO precursor solution can lead to improved photovoltaic performance and stability. To establish that these BZA derivatives result in surface functionalization of the resultant ZnO layers, surface sensitive XPS chemical analysis was performed. Figure 4 presents the XPS survey scans of the modified ZnO samples, with the corresponding surface chemical composition values (atomic % (at. %) by element) summarized in Table 2. All samples exhibit characteristic peaks of zinc (Zn 2p$_{3/2}$, Zn 2p$_{1/2}$), oxygen (O 1s) and carbon (C 1s), which was quantified to assess the chemical composition of each sample. For the ZnO reference sample, the Zn and O content are expected to originate mostly from the metal oxide; atomic concentrations of 34.3 at. % for Zn and 41.4 at. % for O are recorded. It can be seen that the atomic ratio of Zn to O in this sample is not exactly 1:1 with an excess of O. This observation could be attributed to either additional loosely bound oxygen on the surface of ZnO nanocrystals [35], or due to the oxygen present on the surface in the form of adventitious carbon which often incorporates C-O-C and C=O species. Indeed, the observed carbon content (ca. 19.5 at. %) for the ZnO sample is expected to be due to adventitious carbon. The small amount of Si detected (ca. 4.9 at. %) originates from the silicon substrate. For the PEIE-coated ZnO sample, a decrease in Zn and O content along with an increase in C and N content is seen. This result is consistent with what is expected for a thin surface PEIE layer that contains C and N, but is not thick enough to attenuate Zn and O signal from the underlying ZnO layer. For the BZA-modified PEIE sample, a smaller decrease in Zn and O content is observed, along with an increase in carbon content from 19.5 at. % to 40.3 at. % carbon. For both CBA and HBA, which have unique chemical markers (chlorine for CBA and nitrogen for HBA), strong evidence is provided for the presence of these molecules on the surface by the detection of Cl and N in these samples, respectively.

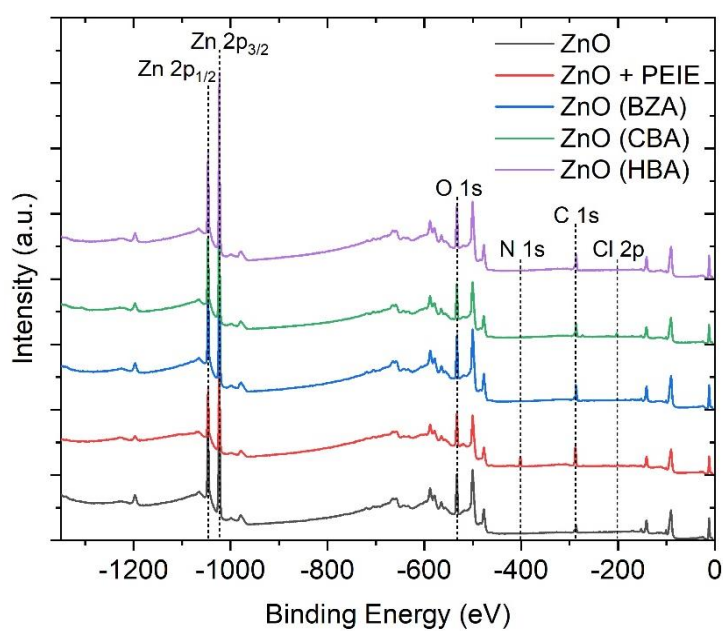

Figure 4. XPS survey scans of ZnO layers prepared with different surface modifiers. The scans have been offset for clarity.

Table 2. Summary of surface chemical composition derived from the XPS data.

| Sample | Zn [at. %] | O [at. %] | C [at. %] | Si [at. %] | N [at. %] | Cl [at. %] |
|---|---|---|---|---|---|---|
| ZnO (Reference) | 34.3 | 41.4 | 19.5 | 4.9 | — | — |
| ZnO + PEIE | 17.2 | 30.7 | 40.3 | — | 11.8 | — |
| ZnO + BZA | 30.2 | 39.7 | 30.1 | — | — | — |
| ZnO + HBA | 28.1 | 37.8 | 32.6 | — | 1.5 | — |
| ZnO + CBA | 29.0 | 35.5 | 31.1 | — | — | 4.4 |

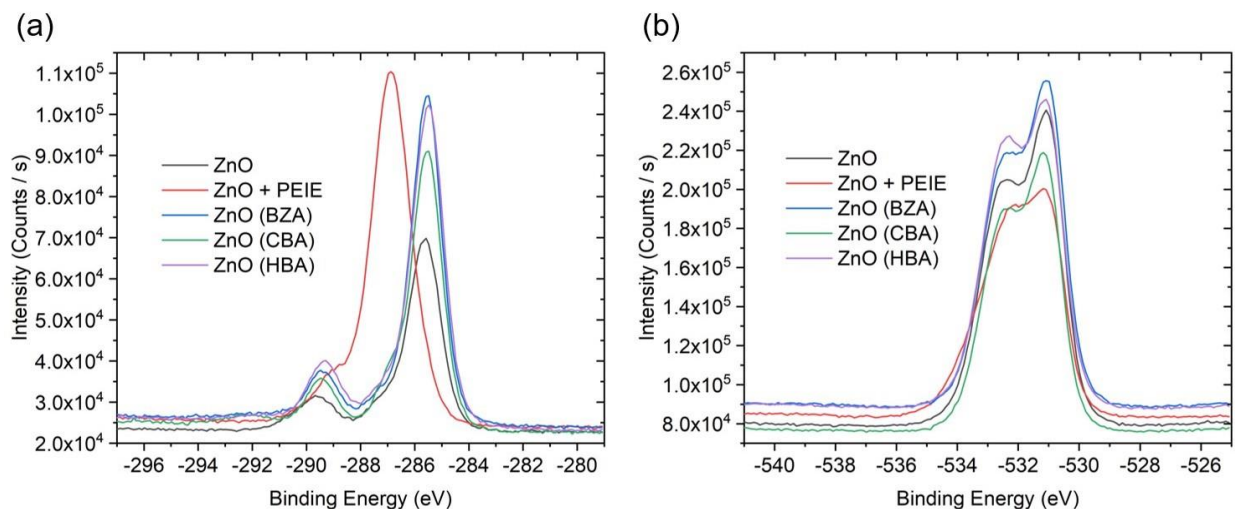

Figure 5. High-resolution XPS scans taken of the (a) C 1s and (b) O 1s peaks.

High resolution XPS scans were also performed covering the C 1s and O 1s peaks, see Figure 5, with the corresponding peak fitting of each sample shown in the Supplementary Material. As shown in Figure 5(a), the PEIE-treated sample appears distinct to the other samples in the XPS scans of the C 1s peaks, characterized by a large peak at around 287 eV, while the main peaks of the other samples are all located around 285 eV. Such change in the position of the main peak in the spectrum can be explained in terms of the distinct chemistry of PEIE, which is a saturated compound compared to the BZA derivatives that are unsaturated compounds (see also Supplementary Material) [36]. For the BZA modified ZnO samples, the main peaks located at about 285 eV can be attributed to aromatic carbon atoms in the benzene ring (refer to Tables S3 & S4 in the Supplementary Material). This peak is highest in the BZA modified sample, closely followed by the HBA modified sample. The ZnO reference sample has the lowest signal with spectral signature similar to the BZA derivatives but also consistent with adventitious carbon. For high resolution XPS scans for the O 1s peak, Figure 5(b), there is not as much variation in the O 1s peak shape (see also corresponding peak fitting in the Supplementary Material). The biggest difference is seen for the PEIE sample which has the lowest signal intensity along with a change in the relative height of the two O 1s peaks seen in this energy range. Oxygen is only present in the form of hydroxyl bonds in PEIE, while for the BZA derivatives oxygen is also present in the form of carbonyl groups. Further detailed analysis of the XPS results can be found in the Supplementary Material.

*2.4.2. Near-Edge X-ray Absorption Fine-Structure (NEXAFS) Spectroscopy Analysis*

As XPS struggles to differentiate between the carbon present in adventitious carbon and that in the BZA derivatives, surface sensitive NEXFAS spectroscopy was also performed, see Figure 6. NEXAFS spectroscopy probes transitions from a core level (e.g. the C 1s state for the carbon K-edge data in Figure 6) to antibonding molecular orbitals. Like XPS, different samples exhibit different NEXAFS spectra owing to differences in surface chemistry. The sensitivity of NEXAFS spectroscopy to the chemistry of anti-bonding orbitals involved in NEXAFS transitions can provide superior chemical sensitivity compared to XPS. Since molecular bonding is closely related to the energy and spatial distribution of the empty electron energy levels, a unique spectroscopic fingerprint can be assigned to each specific chemical moiety or group for the identification of different conjugated materials exhibiting different NEXAFS spectra [37, 38]. Figure 6 shows the C 1s K-edge NEXAFS spectra, with the main peaks labelled with peak assignment provided in Table 3. Clear differences in surface chemistry between the ZnO reference and ZnO samples modified with benzoic acid derivatives is evidenced. Peak 1 at 285 eV in particular is associated with C 1s → π* (C=C) transitions and is prominent in the NEXAFS spectra of the BZA, CBA and HBA modified samples. The prominent nature of this peak for the BZA, CBA and HBA modified samples is consistent with these molecules containing aromatic rings, with the NEXAFS data providing strong evidence for the presence of these molecules on the surface distinct to adventitious carbon. The adventitious carbon instead has a prominent peak at 288.5 eV (Peak 4) consistent with carbonyl units. PEIE also has a distinct NEXAFS spectrum with a prominent peak at 293 eV (Peak 6) associated with C 1s → σ* (C–C) transition, i.e. saturated carbon [37, 38]. More details on the peak assignments and energy positions of the K-edge NEXAFS spectra of carbon are presented in the Supplementary Material.

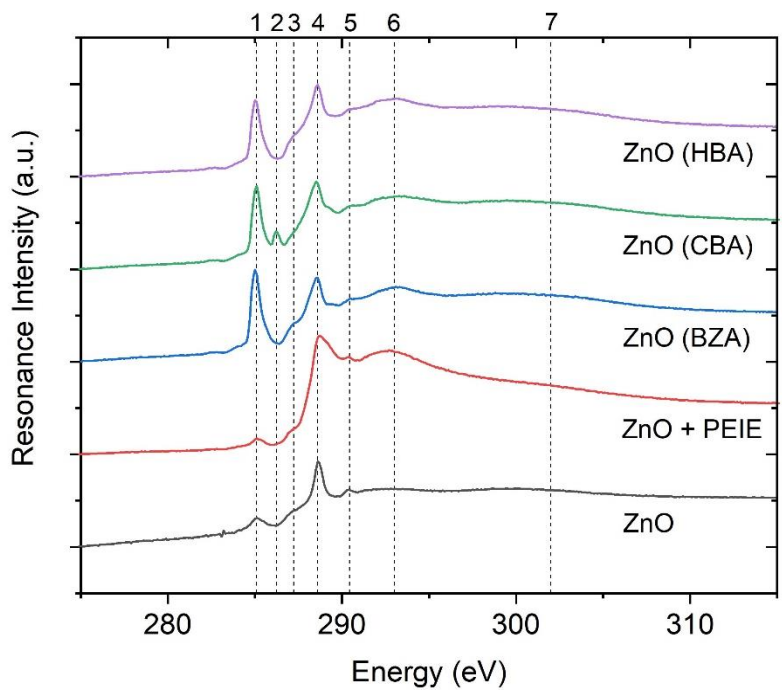

Figure 6. C 1s K-edge NEXAFS spectra of the different ZnO samples with peak identification. Data was acquired at an angle of incidence of 55° X-ray incidence, with the spectra offset for clarity.

Table 3. Peak assignments and energy positions for C 1s K-edge NEXAFS spectra.

| Peak No. | Energy Position [eV] | Peak Notation | Core Level Excitation |
|---|---|---|---|
| 1 | ~ 285.0 | C 1s → π* (C=C) | C-H / C=C |
| 2 | ~ 286.1 | C 1s → π* (C=C) | C-Cl |
| 3 | ~ 287.1 | (Uncertain) | (Uncertain) |
| 4 | ~ 288.5 | C 1s → π* (C=O) | C=O bonded to C-H / C=C |
|   |   | C 1s → σ* (C–H) | Two types of C-H bonds on the benzene rings |
| 5 | ~ 290.5 | C 1s → σ* (C–H) | C–O |
| 6 | ~ 293.0 | C 1s → σ* (C–C) | C-C (C-COOH), C-R (C-Cl or C-N) |
|   |   | C 1s → σ* (C–O) | C-O (C-OH), C-R (C-Cl or C-N) |
| 7 | ~ 302.0 | C 1s → σ* (C=C) | C=C |

By acquiring NEXAFS spectra as a function of different X-ray angles of incidence, the average molecular orientation of molecular units such as the aromatic rings in the BZA derivatives can be probed. Angle-resolved NEXAFS spectra are provided in the Supplementary Material along with further details of the relevant analysis. It was found that for all BZA derivatives, similar tilt angles of ~ 57° to 58° were determined, which corresponds to the angle between the normal to the plane of the aromatic ring and the normal to the substrate. Specifically, average tilt angles of 57.1° for BZA, 57.0° for HBA, and 58.5° for CBA were determined. These results indicate a slight preferential "edge-on" orientation of the molecules. An edge-on orientation can be understood by the tendency for BZA derivatives to anchor to the ZnO surface via the carboxylic acid groups rather than lying flat.

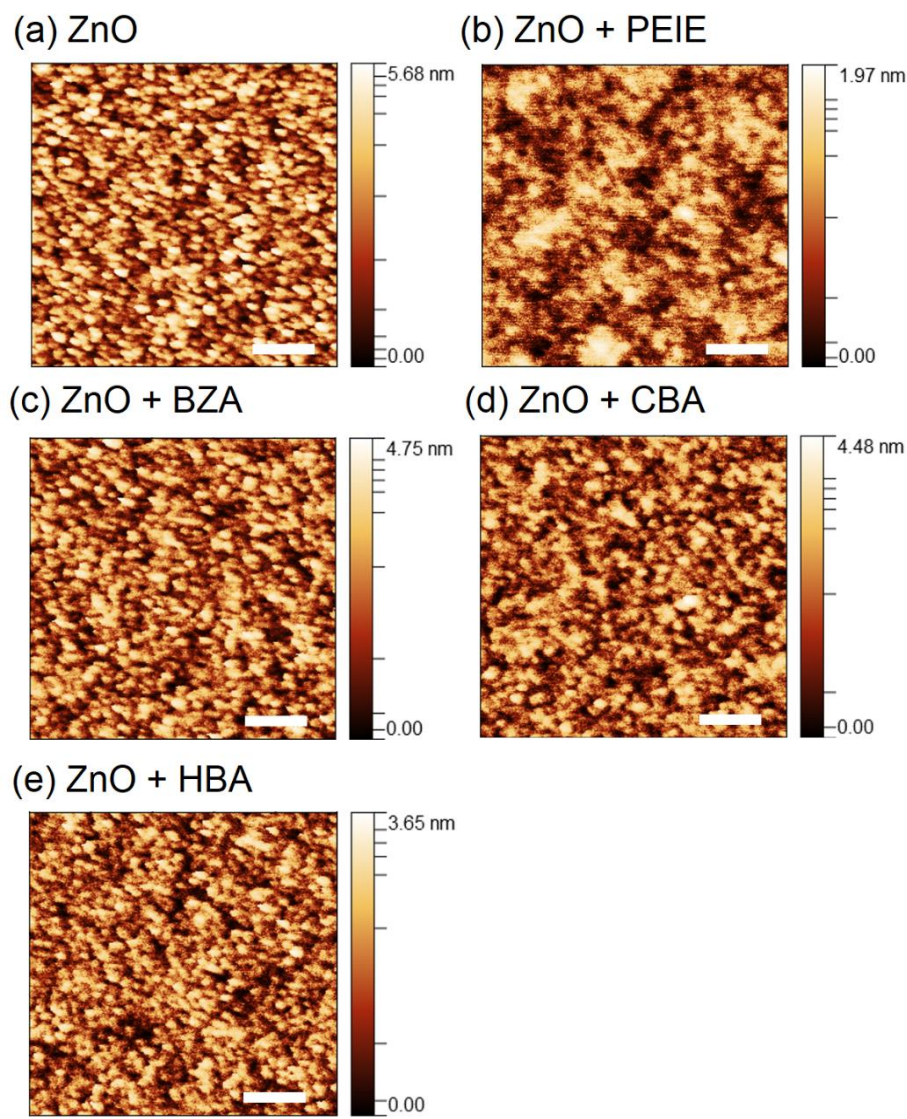

Figure 7. 500 nm by 500 nm AFM height images of (a) ZnO reference; (b) ZnO + PEIE; (c) ZnO + BZA; (d) ZnO + CBA; (e) ZnO + HBA. The scale bar is 100 nm.

*2.4.3. Surface Property Analysis*

Atomic force microscopy (AFM) was performed to investigate the topography of the modified ZnO layers, with the AFM height images shown in Figure 7, and corresponding root-mean-square (RMS) roughness values summarized in Table 4. The surface RMS roughness of ZnO sample decreases significantly after the PEIE layer is applied, from about 0.7 nm down to only about 0.2 nm. This observation is consistent with the PEIE layer coating on top of ZnO surface providing a relatively smooth surface. In the case of the ZnO samples modified with the BZA derivatives, it can be observed that the surface RMS roughness is also significantly smaller than that of the untreated reference ZnO samples. However, the underlying grain structure of the ZnO layer is still clearly visible, indicating that the BZA layers are very thin. The incorporation of the BZA derivatives into the precursor solution also seems to affect the growth of ZnO nanoparticles during film preparation with the grains of ZnO in the BZA, CBA and HBA modified samples appear smaller than that of untreated ZnO. This observation suggests that the BZA derivatives also inhibit the growth of ZnO crystals with HBA having the most prominent effect, which is confirmed by GIWAXS results as will be presented below.

Table 4. Root-mean-square (RMS) roughness, water contact angle and work function for various ZnO-based ETL surfaces.

| Sample | RMS Roughness [nm] | Water Contact Angle [°] | Work Function [eV] |
|---|---|---|---|
| ZnO (Reference) | 0.71 (± 0.06) | 43 (± 2) | 4.42 (± 0.04) |
| ZnO + PEIE | 0.20 (± 0.03) | 19 (± 1) | 4.20 (± 0.06) |
| ZnO + BZA | 0.46 (± 0.04) | 62 (± 4) | 4.29 (± 0.03) |
| ZnO + CBA | 0.38 (± 0.05) | 76 (± 4) | - |
| ZnO + HBA | 0.36 (± 0.04) | 63 (± 2) | - |

As a useful supplement, water contact angle analysis of each sample was performed to better understand how the various modification treatments affect the surface properties of the ZnO layer. Photos of the water contact angle measurements are included in the Supplementary Material, with the derived water contact angles shown in Table 4. Without any modification treatment, the reference ZnO sample has a water contact angle of 43°. When coated with PEIE, the water contact angle reduces significantly to 19°, that is, the sample surface becomes much more hydrophilic. In contrast, treatment of ZnO with any of the BZA derivatives produces a significantly more hydrophobic sample surface, with water contact angles of 62° for BZA modified ZnO, 63° for HBA modified ZnO, and 76° for CBA modified ZnO. The decrease in water contact angle after coating with PEIE is not surprising, as PEIE is a polymeric material with a high content of hydrophilic amine and hydroxyl groups. The increased water contact angle for modification with the BZA derivatives can be associated with the inherent hydrophobic nature of the benzene ring, with the hydrophilic carboxylic acid end group expected to be anchored to the ZnO surface. Furthermore, in the case of the CBA modified ZnO, which has the highest water contact angle, the additional increase in the angle value is likely due to the presence of the chlorine group, which adds further hydrophobicity to the sample surface [39]. In contrast, the presence of the hydrazine group in HBA modified ZnO does not appear to have a significant impact on the surface wetting properties of the sample, as there is no noticeable difference compared to the original BZA modified ZnO sample. As well as supporting the conclusion that the BZA derivatives are surface functionalizing the ZnO layer, the water contact angle results also provide supporting evidence that these derivatives are anchoring to the ZnO via the hydrophilic carboxylic acid groups.

Photoelectron spectroscopy in air (PESA) was also performed to characterize the work function (WF) of the ETL layers, see Table 4 (example PESA data are included in the Supplementary Material). The WF value of the original reference ZnO sample was measured to be 4.42 ± 0.04 eV. Upon modification with PEIE the WF decreases to 4.20 ± 0.06 eV consistent with previous results [23]. For the BZA-modified ZnO layer, the WF is measured to be 4.29 ± 0.03 eV; a decrease in WF but not as pronounced compared to that of PEIE. For CBA and HBA samples a clear PESA signal could not be obtained meaning a reliable WF measurement could not be made for these samples. It is clear that there was a shift in the WFs of these three key samples, confirming that PEIE and BZA modify the WF of the ZnO surface.

*2.4.4. Grazing-Incidence Wide-Angle X-ray Scattering (GIWAXS) Analysis*

GIWAXS measurements were performed to investigate the microstructure of the ZnO layers prepared with different treatments. All samples showing isotropic scattering profiles as shown in the two-dimensional GIWAXS patterns in the Supplementary Material. Figure 8 presents the corresponding radially averaged 1D scattering profiles. Three features (peaks) can be identified between 2 and 2.5 Å$^{-1}$ which can be indexed to (111) peak of the Si substrate [40], and (100) and (002) peaks ZnO [41]. ZnO scattering peaks are observed to be broader in ZnO samples modified with BZA derivatives. Scherrer analysis has been performed to extract the coherence length of ZnO based on the width of (100) ZnO peak enabled by peak fitting with the results summarized in Table 5, with further details on peak fitting provided in the Supplementary Material. A significant difference in the coherence lengths of ZnO is observed for different samples, with the samples modified with BZA derivatives in general having lower coherence lengths. Specifically, the coherence length decreases from 44 Å for the ZnO reference sample to 33 Å and 21 Å for the BZA and HBA modified samples, respectively. On the other hand, coherence length does not decrease significantly for the CBA sample, at 40 Å. PEIE-coated ZnO sample exhibit a coherence length of 42 Å, similar to the reference ZnO sample. In combination with the AFM results, the GIWAXS results provide further evidence that inclusion of BZA derivatives may modulate ZnO crystal growth during preparation.

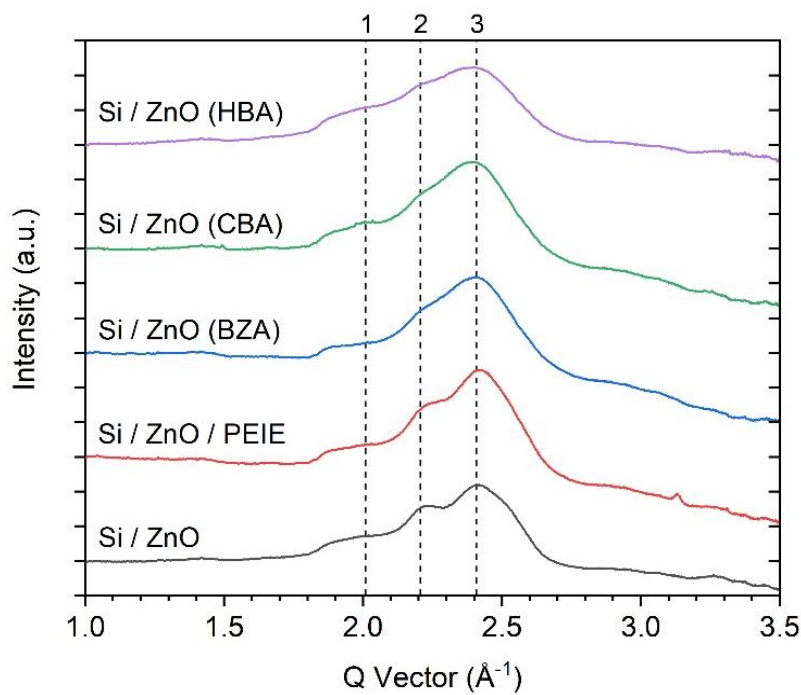

Figure 8. Radially integrated one-dimensional GIWAXS profiles of the ETLs. The profiles were offset in intensity scale for clarity.

Table 5. Summary of diffraction peak assignments and peak fitting results.

| Sample | Peak 1 Position (Si (111)) [Å$^{-1}$] | Peak 2 Position (ZnO (100)) [Å$^{-1}$] | Coherence Length (ZnO (100)) [Å] | Peak 3 Position (ZnO (002)) [Å$^{-1}$] |
|---|---|---|---|---|
| ZnO (Ref.) | 2.03 ± 0.01 | 2.21 ± 0.01 | 44 ± 3 | 2.43 ± 0.01 |
| ZnO + PEIE | 2.03 ± 0.01 | 2.21 ± 0.01 | 42 ± 3 | 2.43 ± 0.01 |
| ZnO + BZA | 1.99 ± 0.01 | 2.19 ± 0.01 | 33 ± 2) | 2.41 ± 0.01 |
| ZnO + CBA | 2.00 ± 0.01 | 2.20 ± 0.01 | 40 ± 3 | 2.40 ± 0.01 |
| ZnO + HBA | 1.95 ± 0.01 | 2.19 ± 0.01 | 21 ± 1 | 2.43 ± 0.01 |

In addition to GIWAXS analysis of the ZnO layers, GIWAXS analysis was also performed on films of PTQ10:ITIC-Br active layer deposited on top of different ETLs in order to assess whether the different interlayer treatments influence the resulting morphology of the BHJ layer. In general, similar BHJ morphology is seen for the different ETL treatments indicating that modification of the ETL preparation process does not have a significant effect on the resulting active layer morphology. The detailed analysis and discussion of the GIWAXS results on the active layer morphology are included in the Supplementary Material.

**2.5 Discussion**

Our results demonstrate that the functionalization of ZnO with BZA derivatives using this one-step approach leads to improved photovoltaic performance and improved operational stability relative to ZnO or ZnO/PEIE ETLs. The surface analysis and microstructural studies confirmed that the addition of BZA derivatives into the ZnO precursor solution leads to a modification of the ZnO layer properties. In particular, from XPS and NEXAFS analysis provided direct evidence of the presence of the BZA derivatives on the surface of the ZnO surface. While the exact coverage or thickness of the BZA molecules on the ZnO is hard to quantify, it is likely to be a very thin layer and certainly thinner than the PEIE layer based on the attenuation of the Zn signal measured by XPS and AFM results. Changes in surface chemistry were also confirmed by water contact angle analysis, with the ZnO layer becoming more hydrophobic. This change in surface energy could be one reason why the stability of the solar cells is improved. The hydrophobic properties of ZnO can trap water, which is a source for the formation of hydroxyl radicals. The formation of a thin layer of BZA derivative on top of ZnO avoids direct contact between ZnO and ITIC-Br which may help mitigate the photochemical reaction of the vinyl group in ITIC-Br [42]. The relatively poor stability of devices containing PEIE can also be attributed to nucleophilic attack from amine groups in PEIE on carbonyl groups in ITIC-Br [43] which is avoided when using a BZA derivative instead. PESA showed that that BZA treatment can lower the work function of the ZnO interlayer, similar to the effect of PEIE. This change in work function may be related to the establishment on an interface dipole with the BZA molecules expected to anchor to the ZnO surface via the carboxylic acid functionality. The increase in water contact angle for the BZA samples is also consistent with this molecular orientation since the aromatic side of the molecule is more hydrophobic than the carboxylic side of the molecule. The improvement in device efficiency can be rationalized by the passivation of traps and the creation of interfacial dipoles, supported by previous works on the modification of ZnO by carboxylic acid functionalized molecules in solar cells [27, 28, 44]. The novelty of this work, however, is that these benefits can be realized simply by adding the BZA molecules directly to the ZnO precursor solution without the need for an additional coating or treatment step. Although the addition of BZA derivatives to the precursor solution seems to reduce ZnO grain size, surprisingly this does not have a negative effect on either device performance or device stability. Indeed, superior device performance and stability can be achieved with respect to coating with PEIE that preserves the microstructure of the underlying ZnO layer. The reason why the HBA derivative gives the best performance in terms of PCE and stability is not clear. While HBA treatment results in the smallest ZnO grain size from GIWAXS analysis, the water contact angle is not the highest (the CBA sample has the highest), being similar to that of BZA-treated ZnO. The XPS analysis suggests that the HBA-treated sample may have the highest surface coverage among all the different BZA molecules based on the decrease in Zn at. % and increase in C at. %; certainly, the degree of coverage is hard to control and one may expect ZnO layers with a complete monolayer coverage expected to result in the best photovoltaic properties. While we have investigated 3 different BZA derivatives here, there are many other BZA derivatives or indeed other surface modifiers [27, 28, 44-46] that could be applied using this one-step approach.

## 3. Conclusion

We have presented a one-step method for the preparation of BZA modified ZnO ETLs through the addition of BZA molecules directly into the ZnO sol-gel precursor solution. This approach was demonstrated for three different BZA derivatives resulting in improved photovoltaic performance and improved operational lifetime with respect to reference ZnO layers. Furthermore, the performance and stability of BZA-modified devices was found to be as good as or better than devices prepared with standard PEIE modification. Through the combination of surface and microstructural analysis, these BZA derivatives were shown to be present on the surface of the ZnO layer after processing, resulting in an increase in water contact angle and decrease in work function (at least for the case of BZA). The addition of BZA derivatives also affects the growth and crystallization of ZnO, though this does not appear to have a negative effect on device performance or device stability. The ability to functionalize metal oxide layers through this simple one-step processing step provides a way to simplify device manufacture.

## 4. Experimental Section
### 4.1 Materials

For ETL preparation, zinc acetate dihydrate (Lot No.: MKCD3805), 2-methoxyethanol (Lot No.: SHBL0639) and ethanolamine (Lot No.: SHBG9313V) were all purchased from Sigma-Aldrich Pty Ltd. For ETL modifiers, PEIE (Lot No.: 04814BGV), BZA (Lot No.: MKCG6487), MBA (Lot No.: 32096DM), HBA (Lot No.: 05201MH) and CBA (Lot No.: MKBQ4957V) were also purchased from Sigma-Aldrich Pty Ltd. For active layer preparation, PTQ10 (Lot No.: SX9035CH) was purchased from 1-Material Inc. with a MW and Đ of 100,000 g/mol and 2.5, respectively. ITIC-Br was synthesized based on previous reports [47, 48], with further details provided in the Supplementary Material.

### 4.2 Materials Preparation

For ZnO precursor solution, the molarity was set at 0.1 M, so that 19 mg of zinc acetate dihydrate precursor was first dissolved in 10 mL of 2-methoxyethanol as solvent, with 60 μL of ethanolamine added as stabilizer. Vigorous stirring of the solution at 60 °C (on a hot plate) for 20 hours was then performed for adequate hydrolysis reaction. Afterwards, the solution was filtered and then used for the desired sol-gel solution [12]. With the same molarity of the precursor (i.e. 0.1 M), the preparation of the BZA modified ZnO solution was very similar to that of the pristine one, where the only difference is that this process involves adding 5 wt.% (~ 11.0 mg) of BZA into the 10 mL zinc acetate dihydrate precursor solution prior to the stirring step as a modifier. For BZA derivatives investigated (i.e. HBA and CBA), the molarities were maintained at the same 0.01 M as for BZA for the sake of comparison. Thus, according to the MW of each derivative (see Figure 1), the amount of HBA and CBA added were set at 13.7 mg and 14.1 mg, respectively. For the active layer mixed solution, a 1:1 weight ratio of PTQ10 and ITIC-Br was dissolved in a nitrogen glovebox in anhydrous chlorobenzene. The resulting solution was then heated on a hot plate in the glovebox overnight before use to ensure that both PTQ10 and ITIC-Br were fully dissolved.

## 4.3 Device Fabrication

All devices were fabricated with an inverted single-junction structure consisting of glass (substrate) / ITO (cathode) / ETL (based on ZnO) / PTQ10:ITIC-Br (active layer) / $MoO_3$ (hole transporting layer, HTL) / Ag (anode). ITO glass substrates were first cleaned in an ultrasonic bath in acetone and then IPA (*isopropanol*) for 10 mins each. Next, substrates were dried with a nitrogen gun, followed by another 10 min of oxygen plasma cleaning. For ZnO ETL (both pristine and BZA or its derivatives modified), about 40 μL of a 0.1 M precursor solution was then dispensed onto a cleaned ITO glass via pipette and spin coated at 3000 rpm for 30 s, and subsequently annealed on a hot plate at 200 ℃ for 30 mins to remove unwanted substances and form a thin conducting metal oxide layer. For some devices, a layer of PEIE (40 μL) was also spin coated atop the ZnO layer at 5000 rpm for 20 s, and annealed on a hot plate at 110 ℃ for 15 mins. After that, the active layer blend of PTQ10 and ITIC-Br was spin coated on the top of each device at 1000 rpm for 2 mins. This step and subsequent steps were performed in a nitrogen glove box to exclude oxygen and water from the device making process. The last two layers of $MoO_3$ HTL and Ag anode were thermally deposited in turn in a vacuum evaporator at around $10^{-7}$ to $10^{-6}$ mbar, in which the samples were mounted on a holder and covered with a shadow mask to determine the evaporated locations and areas, namely the active area (size: ~ 0.15 × 0.3 $cm^2$) of 8 pixels on each device. Finally, after assembling with lead frames, the devices were encapsulated with epoxy resin (Devcon 2-Ton, Lot No.: 101232) and glass slides, which can be taken out of the glove box for further testing after curing for more than 12 hours.

## 4.4 Device Characterization

For light JV measurements, a model SS50AAA 150-Watt Solar Simulator from Photo Emission Tech. was used to simulate sunlight of an AM1.5G radiation spectrum with 100 mW/cm$^2$ irradiance, where intensity of the simulator was calibrated with a silicon reference cell with a KG3 glass filter. For all devices, the current JV characteristics were measured via a Keithley 2635 SourceMeter. The PV parameters, including short-circuit current ($J_{SC}$), open-circuit voltage ($V_{OC}$), fill factor (FF), and power conversion efficiency (PCE), were extracted from the JV curves obtained from these measurements.

### 4.5 X-ray Photoelectron Spectroscopy

XPS measurements were performed on a Thermo Scientific™ Nexsa Surface Analysis System equipped with a hemispherical analyzer and high-efficiency electron lens. Monochromatic Al Kα X-rays (1486.6 eV) at 72 W (6 mA and 12 kV, 400 × 250 μm$^2$ spot) was used on all samples as the incident radiation. Samples were analysed 'As Received' to avoid potential chemical changes to the sample surface post-cleaning. Survey scans collected between −10 eV to 1350 eV were recorded at an analyzer pass energy of 150 eV, a step size of 1.0 eV and a dwell time of 10 ms. High-resolution scans for C 1s, O 1s, Zn 2p and the Zn LMM Auger were obtained with pass energies of 50 eV, step sizes of 0.1 eV and dwell times of 50 ms. The base pressure in the analysis chamber was better than $5.0 \times 10^{-9}$ mbar. Surface charging was removed with the aid of a low-energy dual-beam flood gun. All data were collected and processed using the Thermo Scientific™ Advantage Data System (v5.9931) software; the "Smart" background used for all curve fitting.

### 4.6 Near-Edge X-ray Absorption Fine-Structure (NEXAFS) Spectroscopy

NEXAFS spectroscopy measurements were performed on the Soft X-ray beamline at the Australian Synchrotron using a nearly perfectly linearly polarized X-ray beam. In these measurements, sample films were spin-coated onto a highly doped Si substrate, using the same ETL precursor solutions and under the same processing conditions as for OSC fabrication. For fingerprint analysis, an X-ray incident angle of 55° (known as the "magical angle") was used for all spectra to negate the effects of orientation. Partial electron yield (PEY) data were obtained with a Channeltron detector by recording photoelectrons emitted from the sample with kinetic energy above a certain threshold, determined by a retarding voltage of 251 V. The recorded signals were normalised by the "stable monitor method", with the double normalisation by setting the pre-edge to 0 and the intensity at 320 eV to 1. The collected data were analysed using the Quick AS NEXAFS Tool (QANT) macro package implemented in IgorPro (8.04) [38, 49].

### 4.7 Atomic Force Microscopy (AFM)

AFM measurements were conducted at the Melbourne Centre for Nanofabrication (MCN) using a Bruker Dimension Icon atomic force microscope in ScanAsyst mode (0.8 Hz scan rate). For these measurements, thin films of the same active layer solutions used for the device fabrication were spin-coated onto Si substrates, and the AFM images of film surface topography were then acquired by scanning the films across with a sharp tip. All AFM data were eventually analyzed using Gwyddion.

### 4.8 Water Contact Angle Measurement

Measurements of water contact angles were performed with an OCA 20 contact angle meter, which integrates a video-based optical contact angle measuring and contour analysis system. During the measurements, the SCA 20 Version 2 software package was utilised to control instrument, collect images as well as process data.

### 4.9 Photoelectron Spectroscopy in Air (PESA)

PESA measurements were carried out using a Riken Keiki AC-2 spectrometer with a light intensity of 200 nW and a power number of 0.5. All samples were measured as they were received.

### 4.10 Grazing-Incidence Wide-Angle X-ray Scattering (GIWAXS)

GIWAXS measurements were performed at the SAXS/WAXS beamline at the Australian Synchrotron. The samples used for GIWAXS measurements were prepared on bare Si substrates under the same processing conditions as the OSC fabrication. X-rays with an energy of 15 keV were directed at the sample at an angle of incidence of between $0.05°$ and $0.25°$, with new sample spot used for each measurement to minimize beam-induced damage. The scattering patterns were recorded using a Dectris Pilatus3-2M detector with a pixel size of 0.172 mm × 0.172 mm, located 668 mm downstream of the sample stage, with the exact distance between the sample and the detector calibrated using a silver behenate (AgBeh) standard. The entire path of the X-ray beam was enclosed in vacuum to minimize diffuse air scattering. All data were collected within an exposure time of 1 second, with 3 exposures made per sample taken with different detector positions. These 3 exposures were stitched together to fill the gaps between the detector's module using in-house developed software at the beamline. Data acquired at the critical angle for each sample (identified by the highest scattering intensity) were selected for further analysis [50, 51]. Reduction and analysis of GIWAXS data were performed using a modified version of the Nika macro package implemented in IgorPro (8.04) [52].

**CRediT authorship contribution statement**

**Hao Liu:** Conceptualization, Investigation, Methodology, Writing – original draft, Writing – review & editing. **Chao Wang:** Conceptualization, Investigation, Methodology, Supervision Writing – review & editing. **Wen Liang Tan:** Investigation, Methodology, Writing – review & editing. **Lars Thomsen:** Investigation, Methodology, Writing – review & editing. **Anthony S. R. Chesman:** Investigation, Methodology, Writing – review & editing. **Yvonne Hora:** Investigation, Methodology, Writing – review & editing. **Martyn Jevric:** Resources, Writing – review & editing. **Jonas M. Bjuggren:** Resources, Writing – review & editing. **Mats R. Andersson**: Funding Acquisition, Resources, Supervision, Writing – review & editing. **Yahui Tang:** Investigation, Methodology, Writing – review & editing. **Linjing Tang:** Investigation, Methodology, Writing – review & editing. **Doan Vu:** Conceptualization, Investigation, Methodology, Supervision Writing – review & editing. **Christopher R. McNeill:** Conceptualization, Funding Acquisition, Methodology, Project Administration, Writing – review & editing.

**Declaration of competing interest**

The authors declare that they have no known competing financial interests or personal relationships that could have appeared to influence the work reported in this paper.

**Acknowledgement**

This activity received funding from ARENA as part of ARENA's Research and Development Program—Solar PV Research. The views expressed herein are not necessarily the views of the Australian Government, and the Australian Government does not accept responsibility for any information or advice contained herein. This work was performed in part on the Soft X-ray and SAXS/WAXS beamlines at the Australian Synchrotron, part of ANSTO. This work was also performed in part at the Melbourne Centre for Nanofabrication (MCN) in the Victorian Node of the Australian National Fabrication Facility (ANFF).

**Data availability**

Data will be made available on request.

Supplementary Material.

# One-step Preparation of ZnO Electron Transport Layers Functionalized with Benzoic Acid Derivatives

Hao Liu,[a] Chao Wang,[a] Wen Liang Tan,[a] Lars Thomsen,[b] Anthony S. R. Chesman,[c] Yvonne Hora,[d] Martyn Jevric,[e] Jonas M. Bjuggren,[e] Mats R. Andersson,[e] Yahui Tang,[a] Linjing Tang,[a] Doan Vu,[a] and Christopher R. McNeill[a]*

[a]Department of Materials Science and Engineering, Monash University, Wellington Road, Clayton, Victoria, 3800, Australia
[b]Australian Synchrotron, ANSTO, 800 Blackburn Road, Clayton, Victoria, 3168, Australia.
[c]CSIRO Manufacturing, Ian Wark Laboratories, Clayton, Victoria 3168, Australia
[d]Monash X-ray Platform, Monash University, Wellington Road, Clayton, Victoria, 3800, Australia
[e]Flinders Institute for NanoScale Science and Technology, Flinders University, Adelaide, South Australia, 5042, Australia
*Corresponding author. E-mail: christopher.mcneill@monash.edu

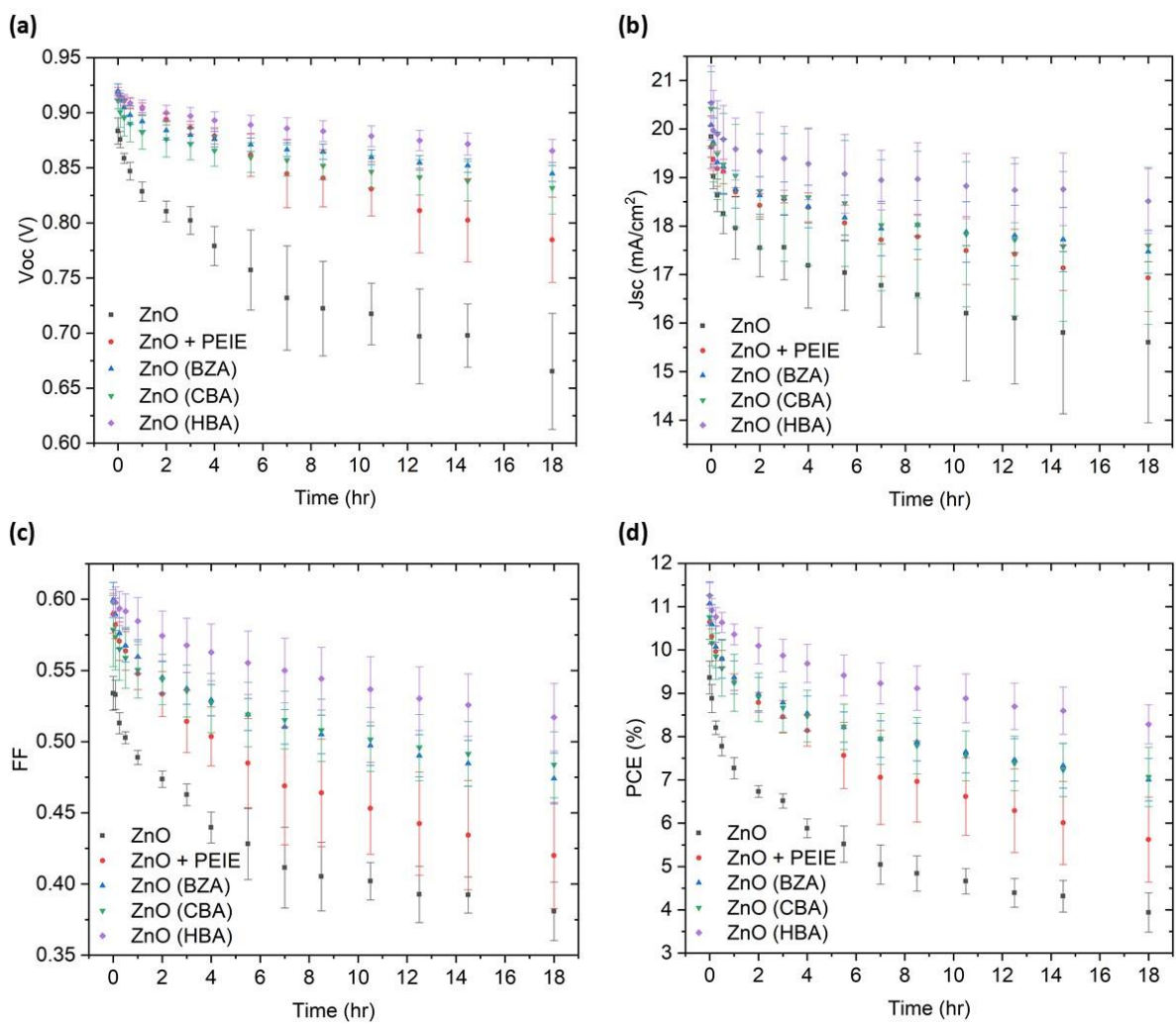

***Figure S1:*** *Original decay patterns of PTQ10 : ITIC-Br based OSCs with various ETLs, where the $V_{OC}$ **(a)**, $J_{SC}$ **(b)**, FF **(c)** and PCE **(d)** are all as a function of exposure time under solar simulator (AM 1.5G, 100 mW/cm$^2$). The mean and standard deviation for each ETL configuration were derived from 5 sets of samples.*

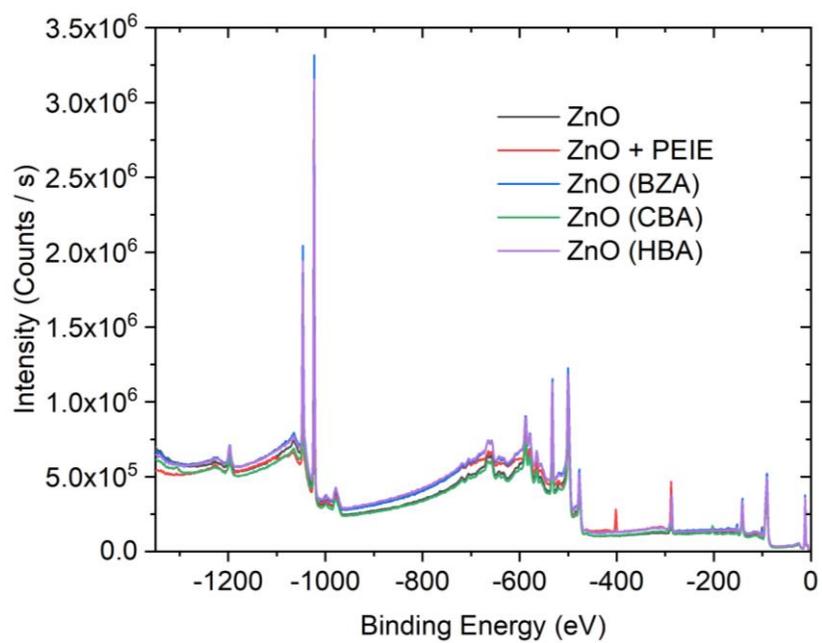

*Figure S2:* Original XPS survey scans of each sample.

| Sample Configuration | Peak Name | Peak BE (eV) | FWHM (eV) | Peak Area (CPS.eV) | Atomic % |
|---|---|---|---|---|---|
| **ZnO (Reference)** | Zn $2p_{3/2}$ | 1023.3 | 2.7 | $7.6 \times 10^6$ | 22.8 |
| | Zn $2p_{1/2}$ | 1046.3 | 2.7 | $3.7 \times 10^6$ | 11.5 |
| | O 1s | 533.0 | 3.3 | $2.1 \times 10^6$ | 41.4 |
| | C 1s | 287.0 | 2.8 | $4.1 \times 10^5$ | 19.5 |
| | Si 2p | 101.1 | 1.5 | $1.0 \times 10^5$ | 4.9 |
| **ZnO (Reference) + PEIE** | Zn $2p_{3/2}$ | 1023.1 | 2.7 | $4.2 \times 10^6$ | 11.3 |
| | Zn $2p_{1/2}$ | 1046.2 | 2.8 | $2.1 \times 10^6$ | 5.9 |
| | O 1s | 532.9 | 3.3 | $1.8 \times 10^6$ | 30.7 |
| | N 1s | 401.6 | 2.2 | $4.3 \times 10^5$ | 11.8 |
| | C 1s | 288.0 | 2.8 | $9.5 \times 10^5$ | 40.3 |
| **ZnO (BZA Modified)** | Zn $2p_{3/2}$ | 1023.2 | 2.7 | $7.5 \times 10^6$ | 19.9 |
| | Zn $2p_{1/2}$ | 1046.0 | 2.7 | $3.7 \times 10^6$ | 10.3 |
| | O 1s | 532.8 | 3.3 | $2.3 \times 10^6$ | 39.7 |
| | C 1s | 286.9 | 1.7 | $7.2 \times 10^5$ | 30.2 |
| **ZnO (HBA Modified)** | Zn $2p_{3/2}$ | 1023.2 | 2.7 | $7.1 \times 10^6$ | 18.6 |
| | Zn $2p_{1/2}$ | 1046.2 | 2.7 | $3.5 \times 10^6$ | 9.5 |
| | O 1s | 532.9 | 3.3 | $2.2 \times 10^6$ | 37.8 |
| | N 1s | 401.1 | 3.4 | $5.4 \times 10^4$ | 1.5 |
| | C 1s | 286.8 | 1.7 | $7.9 \times 10^5$ | 32.6 |
| **ZnO (CBA Modified)** | Zn $2p_{3/2}$ | 1023.3 | 2.6 | $6.5 \times 10^6$ | 19.1 |
| | Zn $2p_{1/2}$ | 1046.4 | 2.2 | $3.2 \times 10^6$ | 9.9 |
| | O 1s | 533.0 | 3.1 | $1.8 \times 10^6$ | 35.5 |
| | C 1s | 287.0 | 2.6 | $6.6 \times 10^5$ | 31.1 |
| | Cl 2p | 202.5 | 2.2 | $2.1 \times 10^5$ | 4.4 |

***Table S1:*** *Peak table of XPS survey scan, where the Peak Area is equivalent to profiles; the FWHM (full-width at half-maximum)) is a key indicator of chemical state change and physical influences.*

| Sample | Name | Peak BE (eV) | FWHM (eV) | Peak Area (CPS.eV) | Atomic % |
|---|---|---|---|---|---|
| ZnO (Reference) | C 1s (the C from adventitious C contamination (C-C)) | 284.8 | 1.4 | 70102.4 | 17.0 |
| | C 1s (the C from adventitious C contamination (O-C=O)) | 288.9 | 1.9 | 16131.5 | 3.9 |
| | C 1s (the C from adventitious C contamination (C-O-C)) | 286.4 | 1.3 | 9757.0 | 2.4 |
| | O 1s (the O from metal oxide (ZnO) lattice) | 530.2 | 1.3 | 191324.6 | 19.1 |
| | O 1s (the O from loosely bound $O_2$ on sample surface; $O^{2-}$ ions in the oxygen deficient regions) | 531.7 | 1.8 | 239342.8 | 24.0 |
| | Zn $2p_{3/2}$ (ZnO) | 1021.5 | 1.8 | 1411462.7 | 21.5 |
| | Zn $2p_{1/2}$ (ZnO) | 1044.5 | 2.1 | 770881.7 | 12.2 |
| ZnO (Reference) + PEIE | C 1s (the C from adventitious C contamination or polymer (PEIE) (C-C)) | 285.5 | 2.0 | 40238.5 | 8.9 |
| | C 1s (the C from adventitious C contamination (O-C=O)) | 288.1 | 2.1 | 25721.2 | 5.7 |
| | C 1s (the C from polymer (PEIE) (C-O or C-N)) | 286.1 | 1.5 | 113635.9 | 25.1 |
| | O 1s (the O from metal oxide (ZnO) lattice) | 530.2 | 1.2 | 114049.2 | 10.4 |
| | O 1s (the O from PEIE (organic C-O)) | 531.4 | 1.1 | 31508.9 | 2.9 |
| | O 1s (the O from loosely bound $O_2$ on sample surface; $O^{2-}$ ions in the oxygen deficient regions) | 531.7 | 2.5 | 209958.8 | 19.2 |
| | N 1s (the polymer (PEIE) N) | 399.9 | 1.8 | 75069.4 | 10.7 |
| | Zn $2p_{3/2}$ (ZnO) | 1021.4 | 1.6 | 786014.1 | 11.0 |
| | Zn $2p_{1/2}$ (ZnO) | 1044.5 | 1.8 | 423477.8 | 6.1 |
| ZnO (BZA Modified) | C 1s (the aromatic (phenyl) C with COOH as nearest neighbour) | 284.8 | 1.9 | 38033.0 | 8.1 |
| | C 1s (the carboxyl or carbonyl C (C=O), or the C from adventitious C contamination (O-C=O)) | 288.6 | 1.3 | 14859.3 | 3.2 |
| | C 1s (the p-p* satellite / shake-up structure) | 290.4 | 3.4 | 3923.6 | 0.8 |
| | C 1s (the five equivalent aromatic ring (phenyl) C), or the C from adventitious C contamination (C-C)) | 284.7 | 1.1 | 73576.8 | 15.6 |
| | C 1s (the carboxyl C (C-O), or the C from adventitious C contamination (C-O-C)) | 286.0 | 1.9 | 10696.4 | 2.3 |
| | O 1s (the O from BZA (organic C-O or C=O) or hydroxyl (OH) groups) | 531.8 | 1.2 | 20388.2 | 1.8 |
| | O 1s (the O from loosely bound $O_2$ on sample surface; $O^{2-}$ ions in the oxygen deficient regions) | 531.6 | 1.9 | 220884.5 | 19.4 |
| | O 1s (the O from metal oxide (ZnO) lattice) | 530.2 | 1.3 | 200836.4 | 17.6 |
| | Zn $2p_{3/2}$ (ZnO) | 1021.5 | 1.8 | 1487761.7 | 19.9 |
| | Zn $2p_{1/2}$ (ZnO) | 1044.5 | 2.1 | 816691.2 | 11.3 |

***Table S2:*** *Peak deconvolution table for XPS narrow scan each sample.*

| Sample | Name | Peak BE (eV) | FWHM (eV) | Peak Area (CPS.eV) | Atomic % |
|---|---|---|---|---|---|
| ZnO (CBA Modified) | C 1s (the five equivalent aromatic ring (phenyl) C), or the C from adventitious C contamination (C-C)) | 284.7 | 1.2 | 81636.7 | 19.8 |
| | C 1s (the carboxyl C (C-O), or the C from adventitious C contamination (C-O-C)), or the aromatic (phenyl) C with Cl as nearest neighbour (C-Cl)) | 286.1 | 1.2 | 13000.3 | 3.2 |
| | C 1s (the carboxyl or carbonyl C (C=O), or the C from adventitious C contamination (O-C=O)) | 288.6 | 1.3 | 15358.2 | 3.7 |
| | C 1s (the p-p* satellite / shake-up structure) | 291.4 | 2.2 | 5191.7 | 1.3 |
| | C 1s (the aromatic (phenyl) C with COOH as nearest neighbour) | 284.8 | 2.6 | 15925.4 | 3.9 |
| | O 1s (the O from CBA (organic C-O or C=O) or hydroxyl (OH) groups) | 531.6 | 1.2 | 29276.7 | 2.9 |
| | O 1s (the O from loosely bound $O_2$ on sample surface; $O^{2-}$ ions in the oxygen deficient regions) | 531.8 | 1.8 | 163494.3 | 16.4 |
| | O 1s (the O from metal oxide (ZnO) lattice) | 530.3 | 1.3 | 174642.0 | 17.5 |
| | Cl $2p_{3/2}$ (organic) | 200.5 | 1.2 | 22557.8 | 1.9 |
| | Cl $2p_{1/2}$ (organic) | 202.1 | 1.2 | 10248.3 | 0.9 |
| | Cl $2p_{1/2}$ (chloride) | 198.6 | 1.6 | 5304.2 | 0.4 |
| | Zn $2p_{3/2}$ (ZnO) | 1021.5 | 1.8 | 1172665.9 | 18.0 |
| | Zn $2p_{1/2}$ (ZnO) | 1044.5 | 2.1 | 639593.3 | 10.1 |
| ZnO (HBA Modified) | C 1s (the carboxyl or carbonyl C (C=O), or the C from adventitious C contamination (O-C=O)) | 288.6 | 1.3 | 20263.4 | 4.3 |
| | C 1s (the aromatic (phenyl) C with COOH as nearest neighbour) | 284.8 | 2.9 | 20158.1 | 4.2 |
| | C 1s (the five equivalent aromatic ring (phenyl) C), or the C from adventitious C contamination (C-C)) | 284.7 | 1.3 | 95874.6 | 20.1 |
| | C 1s (the carboxyl C (C-O), or the C from adventitious C contamination (C-O-C)), or the aromatic (phenyl) C with N as nearest neighbour (C-N)) | 286.1 | 1.8 | 14103.9 | 3.0 |
| | C 1s (the p-p* satellite / shake-up structure) | 290.7 | 2.6 | 2760.3 | 0.6 |
| | O 1s (the O from loosely bound $O_2$ on sample surface; $O^{2-}$ ions in the oxygen deficient regions) | 531.7 | 1.6 | 207709.9 | 18.0 |
| | O 1s (the O from HBA (organic C-O or C=O) or hydroxyl (OH) groups) | 531.5 | 3.4 | 51806.4 | 4.5 |
| | O1s (the O from metal oxide (ZnO) lattice) | 530.2 | 1.3 | 181354.5 | 15.7 |
| | N 1s (the hydrazino group N (HBA)) | 399.3 | 2.5 | 6428.7 | 0.9 |
| | Zn $2p_{3/2}$ (ZnO) | 1021.5 | 1.8 | 1398991.2 | 18.5 |
| | Zn $2p_{1/2}$ (ZnO) | 1044.5 | 2.1 | 762201.8 | 10.4 |

***Table S3:*** *The peak deconvolution table for XPS narrow scan each sample (continued).*

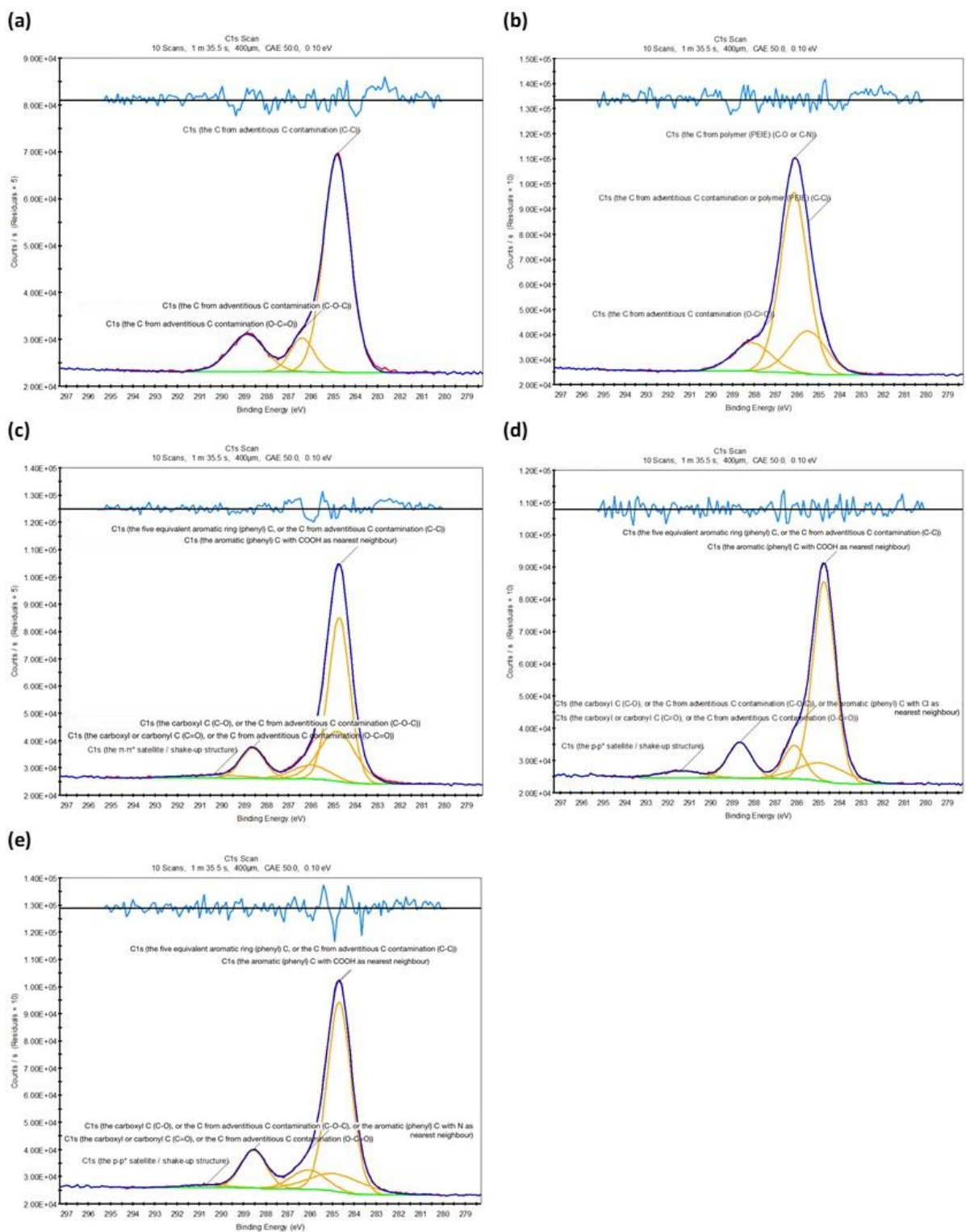

*Figure S3:* XPS narrow scan spectra of C 1s peaks of *(a)* ZnO (Reference); *(b)* ZnO (Reference) + PEIE; *(c)* ZnO (BZA Modified); *(d)* ZnO (CBA Modified); *(e)* ZnO (HBA Modified).

From Figure S3 (a) and Table S2, the C detected for the untreated reference ZnO sample is predominantly from adventitious carbon. Adventitious carbon is also a non-negligible source of the C 1s signal for the C 1s peaks of the other samples.[1] For the samples treated with either

S7

BZA, CBA or HBA (see Figure S3 (c, d, e) and Table S2 & Table S3), the C 1s peak is mainly contributed by carbon atoms at equivalent positions of the aromatic (phenyl) ring (ca. 15.6 at. % for BZA, 19.8 at. % for CBA, 20.0 at. % for HBA), which is reflected in the highest peak centred about 284.7 eV in their spectra. If the aromatic carbon atoms, which have a slightly higher binding energy (ca. 284.8 eV) due to the proximity of the carboxyl group (-COOH), are also considered, the carbon signal provided by the entire benzene ring can be further increased to approximately 24 at. %.[2] This is followed by carbon atoms of the carboxylic acid group attached to the benzene ring (ca. 3.2 at. % for BZA, 3.7 at. % for CBA, 4.3 at. % for HBA) that is embodied in the second highest peak located on the left side of the spectra at around 288.6 eV in the same figures.[1, 2] These atomic ratios can further rise to around 5 to 7 % if not only the carbon-oxygen double bond (C=O) in the carboxyl or carbonyl group is considered, but also the carbon-oxygen single bond (C-O) therein. In fact, it can be found that the above measured relative atomic ratios of carbon atoms from the benzene rings and the carboxyl groups of the BZA derivatives, are not in complete agreement with the compositional stoichiometry of their corresponding chemical formulae shown in Figure 1. Apart from the potential measurement errors, the most likely reason is the superposition of the signal from the adventitious carbon with the C 1s peaks of interest, leading to an overestimation of the intensity of each component peak. This conjecture at least is consistent with the C 1s measurements in the reference ZnO sample that are only associated with adventitious carbon.

Notably, some "shake-up" features, also known as satellites, can also be identified in the narrow scan spectra of the BZA derivative modified samples.[3] These subtle satellites originate due to $\pi \rightarrow \pi^*$ transitions in the molecules, in which the un-localized electrons in highest occupied molecular orbitals (HOMOs) are promoted to lowest unoccupied molecular orbitals (LUMOs).[1, 3] This transition will be more apparent in the NEXAFS analysis.

Regarding the results of the PEIE coated samples, Figure S3 (b) and Table S2, which provide some confirmation that the PEIE covers the entire surface of the samples to a great extent, as over 85% of the carbon detected can be attributed to PEIE, no matter whether the substrate is untreated reference ZnO (about 34.0 at. % of C comes from PEIE), or ZnO that has been modified by BZA first (about 41.2 at. % of C comes from PEIE). This agrees with the PEIE structure formula in Figure 1, in which it can be seen that the most important component of this polymer is carbon.

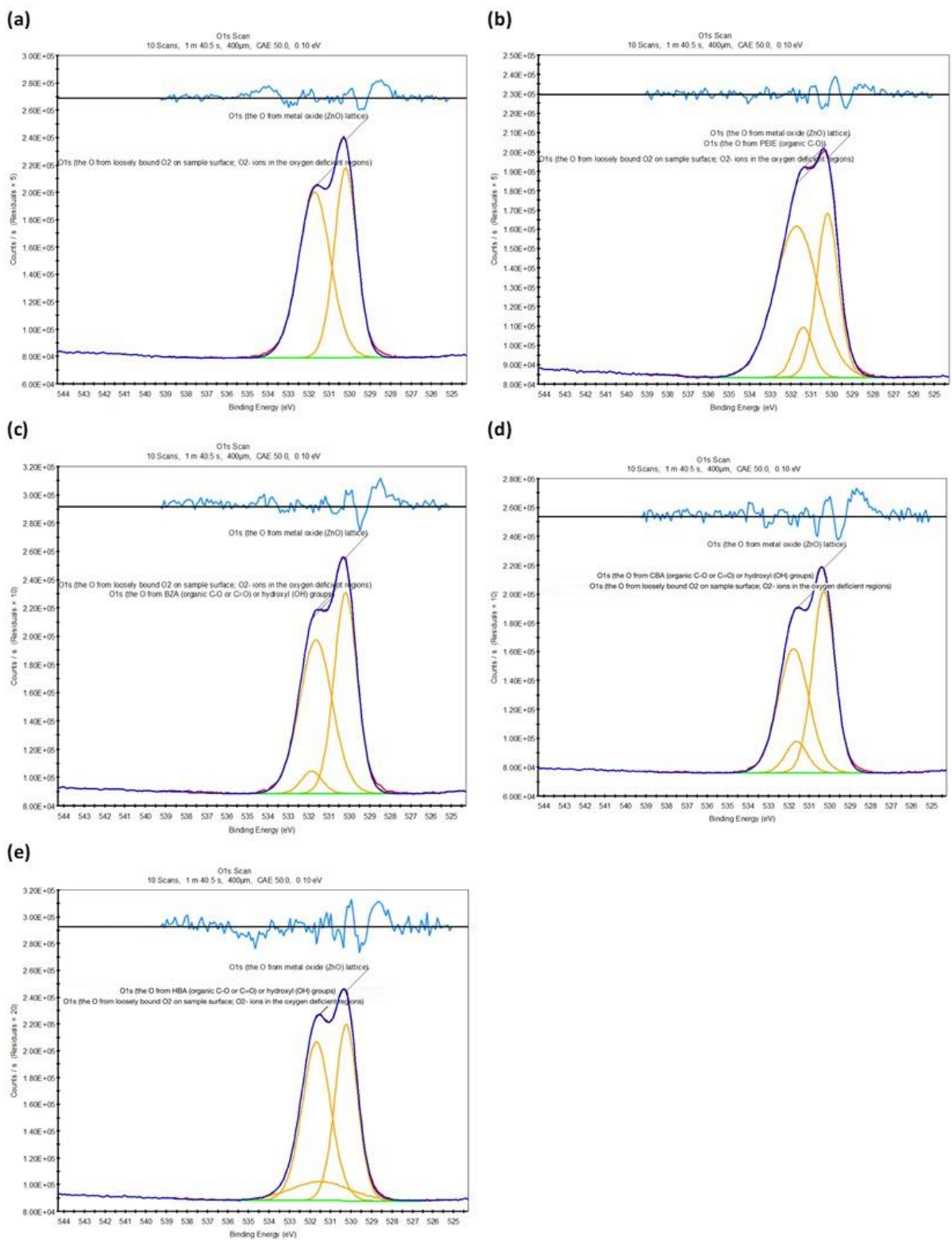

***Figure S4:*** *XPS narrow scan spectra of O 1s peaks of **(a)** ZnO (Reference); **(b)** ZnO (Reference) + PEIE; **(c)** ZnO (BZA Modified); **(d)** ZnO (CBA Modified); **(e)** ZnO (HBA Modified).*

The O 1s peaks of all samples are asymmetric and can be deconvoluted into three peaks with different binding energy except for that of the untreated reference ZnO sample which has only two component peaks, see Figure S4. According to Table S2 and Table S3, the dominant peak which has lowest binding energy of around 530.2 to 530.3 eV is attributed to oxygen (i.e. $O_2^-$ ions) in the ZnO crystal lattice that is surrounded by Zn.[2, 4-6] The other main peak with higher binding energy located at around 531.6 to 531.8 eV is tentatively associated with oxygen from loosely bound $O_2$ on the sample surface; $O_2^-$ ions in the oxygen deficient regions; and chemisorbed oxygen caused by interfacial functional groups like hydroxyl (-OH).[2, 5, 6] The third, minor oxygen peak can be associated with the specific chemistry of the PEIE or BZA-derivatives. The first two components are the main contributors to the O 1s spectrum for each sample herein, while for the BZA derivative modified samples, there may be a relatively weak component peaks between the two as well (see Figure S4 (c, d, e)). These insignificant O 1s peaks should be attributed mainly to the oxygen in the carboxylic acid group of the BZA derivatives (see Figure 1).[1] The reason for such low peak intensities is most likely due to the fact that these modifiers are not covered as a separate layer on the surface of the pristine ZnO sample, but are mixed with the latter. This also partly explains why the O 1s spectrum of the reference ZnO sample, which has only two component peaks on the left and right, is so similar to that of the BZA.

As for the sample to which PEIE was also applied, the main effect is an overall decrease in the intensity of the O 1s peaks, which is particularly pronounced for the component peaks originating from the ZnO lattice (~ 530 eV). The most intuitive reason for this is that the PEIE layer covering the surface of the samples largely prevents the oxygen beneath it from being detected, while contains very little oxygen itself (see Figure 1).

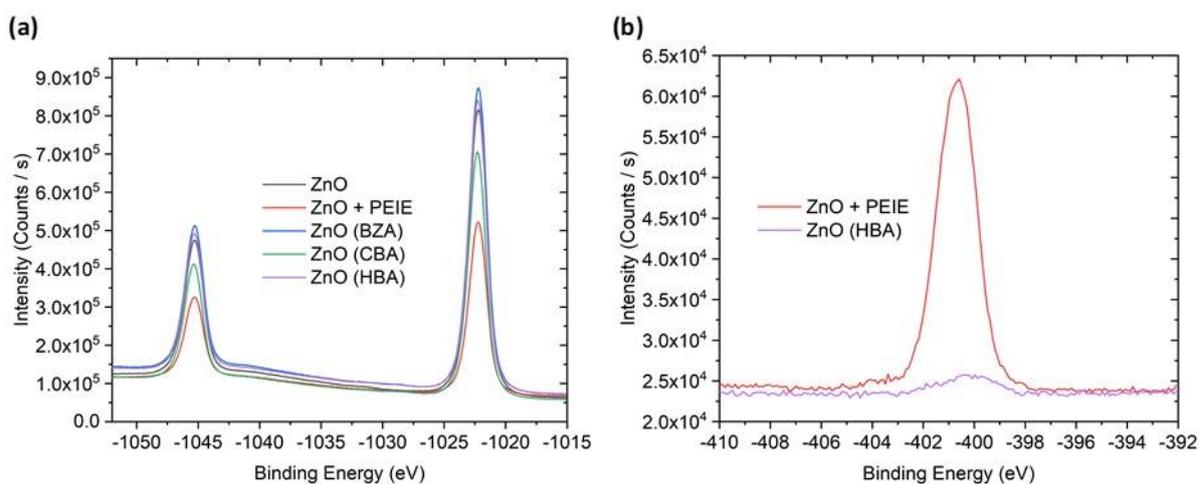

*Figure S5:* XPS narrow scan spectra of *(a)* Zn 2p peaks and *(b)* N 1s peaks of the different samples.

Figure S5 presents the Zn 2p peak and N 1s peak narrow energy scans. Because the only source of Zn is the ZnO substrate itself, the effect of various modifications on the Zn 2p peaks is basically only in the peak intensity. According to Figure S5 (a) and Table S1, the peak intensity of both the left spectral peak (Zn $2p_{1/2}$, ~ 1046 eV) and the right spectral peak (Zn $2p_{3/2}$, ~ 1023 eV) is the largest for the BZA modified sample, followed by the HBA modified sample, while the peak intensity of the CBA modified sample is rather inferior to that of the original reference ZnO sample. Besides, the Zn 2p peak intensities of the PEIE coated sample is significantly lower than those of all the other samples, which again validates the possibility of attenuation of the underlying signal by PEIE. As for the N 1s peaks shown in Figure S5 (b), the presence of PEIE or HBA on the sample surface can be effectively demonstrated, although the associated peak of the latter may be largely negligible because it is not directly applied to the sample surface, or contains less nitrogen itself (see Figure 1).

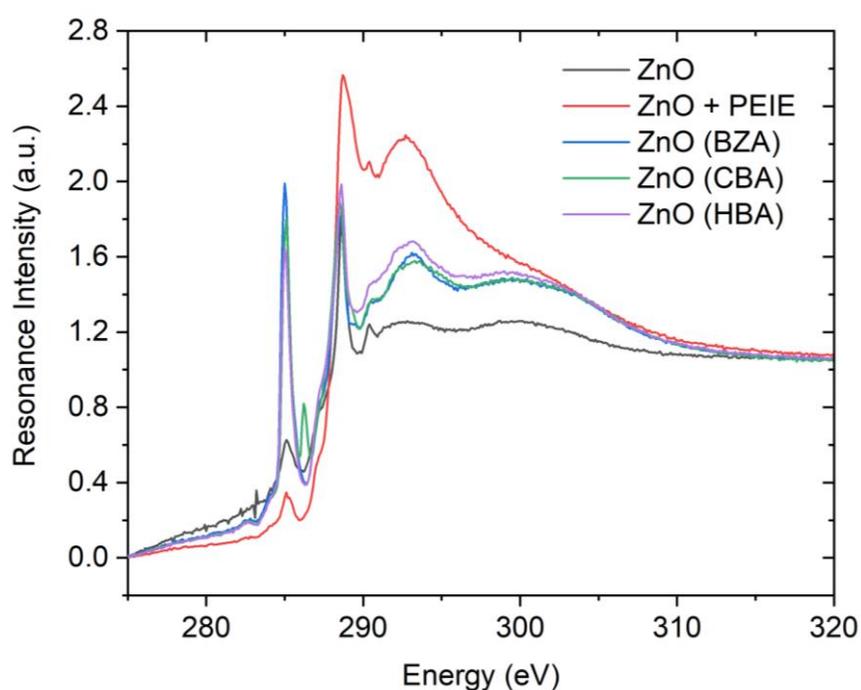

***Figure S6:*** *C 1s K-edge NEXAFS spectra for each spin-coated film sample at 55° X-ray incidence, where all spectra are overlaid.*

| Peak No. | Energy Position (eV) | Peak Notation (w/ Upper Level) | Core Level | Explanation | | |
|---|---|---|---|---|---|---|
| Peak 1 | ~ 285.0 | C 1s → π* (C=C) | Unsubstituted (unsaturated) aromatic carbon atoms (i.e. *C-H* or *C=C*) (The *C=C* designation here is not very accurate, since there is actually no defined double bond in benzene ring.) | The transitions of the unsubstituted ring carbons are grouped in an experimentally unresolved energy range, leading to the 'aromatic peak' at around 285.0 eV. Additionally, it is assumed that the C 1s (*C-COOH*) → π* (*C=C*) transition is located at the high energy tailing edge of the aromatic peak. | The antibonding orbitals associated with different neighbour bonds should be similar in energy. E.g., *C-C* and *C-O* σ* orbitals, or *C=C* and *C=O* π* orbitals, respectively. | The C 1s → π* resonances are derived from the conjugated backbone of the polymers. They are almost identical throughout all the configurations of bare BZA derivatives, since only the backbone structure contains the unsaturated carbon bonds (like *C=C*) required for π-type bonding. Accordingly, for the configurations with PEIE covered, the detected resonances and relative peak intensities are obviously weakened. |
| Peak 2 | ~ 286.1 | C 1s → π* (C=C) | Substituted (unsaturated) aromatic carbon atoms (i.e. *C-Cl*) (The C=C designation here is not very accurate, since there is actually no defined double bond in benzene ring.) | The (energy) position of this peak correlates with the electronegativity (i.e. electron withdrawing effect) of the halogen substituted atom: F (287.1 eV) > O (286.8 eV) > N (286.5 eV) > Cl (286.1 eV) > Br (285.9 eV) > I (285.6 eV). The halogen substituents can be regarded as simple electron withdrawing point charges. | | |
| Peak 3 | ~ 287.1 | N.A. | N.A. | Cannot find the corresponding interpretation in the literature. The signal may come from adventitious carbon or contamination. | | |

*Table S4: Detailed description of the peak assignments and energy positions for Carbon K-edge NEXAFS spectra at 55° X-ray incidence.*

| Peak No. | Energy Position (eV) | Peak Notation (w/ Upper Level) | Core Level | Explanation | | | |
|---|---|---|---|---|---|---|---|
| Peak 4 | ~ 288.5 | C 1s → σ* (*C-C*) | (Saturated) carboxylic (carboxyl functional group) carbon atoms (i.e. *C=O*) bonded to (unsaturated) aromatic carbon (*C-H* or *C=C*) | An aromatic or unsaturated C atom bonded to another C atom has a strong 1s → π* transition at about 285.0 eV.<br>However, as more electron-drawing atoms such as O are added or substituted, the binding energy of C 1s electron increases.<br>This increase in binding energy shifts the 1s → π* transitions of aromatic C (*C-H* or *C=C*) to higher energies, that is, up to 288.5 eV for the two O atoms bonded to saturated C in carboxyl functional groups (*C=O*). | The antibonding orbitals associated with different neighbour bonds should be similar in energy. | The C 1s → π* resonances are derived from the conjugated backbone of the polymers.<br>They are almost identical throughout all the configurations of bare BZA derivatives, since only the backbone structure contains the unsaturated carbon bonds (like *C=C*) required for π-type bonding.<br>Accordingly, for the configurations with PEIE covered, the detected resonances and relative peak intensities are obviously weakened. | A reasonable explanation for the lack of fine structure in the spectra shown here is that the extensive delocalization of the molecular orbitals of these conjugated polymer materials exhibit the electronic structure of a broader band structure. |
| | | C 1s → σ* (*C–H*) | (Unsaturated) sp$^2$ hybridized carbon atoms of benzene rings (i.e. *C-H* (2 types)) | Generally the C 1s → σ* (*C–H*) should be mainly contributed by the side chains of the materials, the side chains that vary between the materials should be apparent in the variation of such resonances.<br>However, in this case, the side chains of all materials are not aliphatic.<br>Thus, it can only be guessed that the 1s → σ* transitions should come from the sp$^2$ hybridized aromatic C.<br>Beside, in the case with PEIE, both the intensity and width of this peak are significantly improved. This should be attributed to the *C-H*, *C-C* and *C-N* bonds in the polymer. | E.g., *C-C* and *C-O* σ* orbitals, or *C=C* and *C=O* π* orbitals, respectively. | | This should be particularly pronounced for amorphous polymers, which show more varied local electronic structure than ordered, i.e., crystalline, polymers, or mixtures of ordered and disordered materials. |
| Peak 5 | ~ 290.5 | C 1s → σ* (*C–H*) | *C–O* | Could be from adventitious carbon or contamination. | | | |

*Table S5: (Continued) detailed description of the peak assignments and energy positions for Carbon K-edge NEXAFS spectra at 55° X-ray incidence.*

S13

| Peak No. | Energy Position (eV) | Peak Notation (w/ Upper Level) | Core Level | Explanation | | |
|---|---|---|---|---|---|---|
| Peak 6 | ~ 293.0 | C 1s → σ* (*C-C*) | (Saturated) sp² hybridized carbon atoms of carboxylic groups (i.e. *C-C* (*C-COOH*), *C-R* (*C-Cl or C-N*)) | In the case of applying PEIE, the intensity of this peak is distinctly improved. This should be attributed to the *C-C* and *C-N* bonds in the polymer. | The antibonding orbitals associated with different neighbour bonds should be similar in energy. E.g., *C-C* and *C-O* σ* orbitals, or *C=C* and *C=O* π* orbitals, respectively. | A reasonable explanation for the lack of fine structure in the spectra shown here is that the extensive delocalization of the molecular orbitals of these conjugated polymer materials exhibit the electronic structure of a broader band structure. This should be particularly pronounced for amorphous polymers, which show more varied local electronic structure than ordered, i.e., crystalline, polymers, or mixtures of ordered and disordered materials. |
| | | C 1s → σ* (*C-O*) | (Saturated) sp² hybridized carbon atoms of carboxylic groups (i.e. *C-O* (*C-OH*), *C-R* (*C-Cl or C-N*)) | Analogously, as more electron-drawing atoms (e.g., O, N, Cl) are added or substituted, the binding energy of C 1s electron increases. This increase in binding energy shifts the 1s → σ* transitions of aromatic C (*C-H*) to higher energies, for example, around 293.0 eV in the case of the two O atoms bonded to saturated C in carboxylic groups (*C-O*). | | |
| Peak 7 | ~ 302.0 | C 1s → σ* (*C=C*) | Unsaturated aromatic carbon atoms (i.e. *C=C*) (The *C=C* designation here is not very accurate, since there is actually no defined double bond in benzene ring.) | In the case of covering with PEIE, the intensity of the peak is attenuated, although it is not otherwise obvious. | | |

***Table S6:*** *(Continued) detailed description of the peak assignments and energy positions for Carbon K-edge NEXAFS spectra at 55° X-ray incidence.*

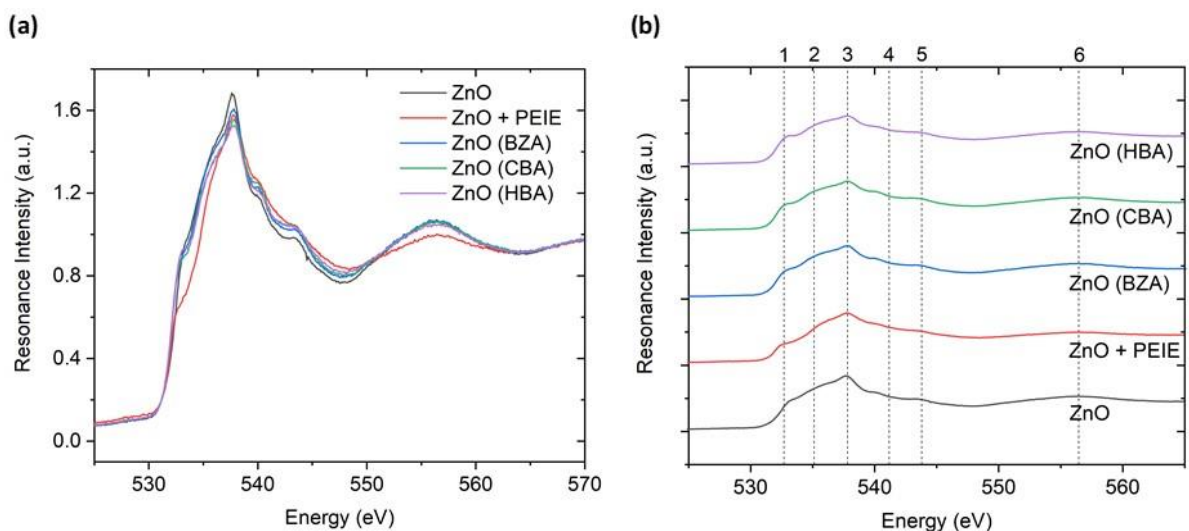

*Figure S7:* O 1s K-edge NEXAFS spectra with peak identification for each spin-coated film sample at 55° X-ray incidence. In part *(a)* all spectra are overlaid, while in part *(b)* the spectra for each sample are offset for clarity.

| Peak No. | Energy Position (eV) | | Explanation |
|---|---|---|---|
| 1 | ~ 532.6 | | O 2p – Zn 3d hybridization (possible) |
| 2 | ~ 535.1 | O 2p – Zn 4s hybridization | N.A. |
| 3 | ~ 537.8 | | Transition to the non-dispersive O 2p states (i.e. $2p_x$ & $2p_{x+y}$) |
| 4 | ~ 541.1 | O 2p – Zn 4p hybridization | N.A. |
| 5 | ~ 543.8 | | N.A. |
| 6 | ~ 556.4 | O 2p – Zn 4d hybridization | |

*Table S7:* Peak assignments and energy positions for O 1s K-edge NEXAFS spectra at 55° X-ray incidence.

O 1s K-edge NEXAFS spectra taken at 55° X-ray incident angle with peak assignments are shown in Figure S7 and Table S7, where the spectral features from the energy range measured can be broadly divided into six peaks. The oxygen K-edge is at around 530 eV, while the peak (or peak-like feature) with the lowest energy position is located at about 532.6 eV, especially for the samples without the covering of PEIE.[7] As the energy of this peak is close to the conduction band minimum, the peak can be attributed to the hybridisation of the O 2p states with the Zn 3d states.[8] The peak with highest resonance intensity at around 537.8 eV (i.e. Peak 3) should be associated with the transition to the non-dispersive O $2p_x$ and O $2p_{x+y}$ states.[8] However, the plausible signal source for the Peak 2 that may exist between these two peaks remains uncertain. Alternatively, it is simply a peak-like feature that is essentially part of adjacent peaks. Thus, if the Peak 1, 2 and 3 are considered as a whole, the spectral features in the energy range of 528 to 539 eV can be ascribed to the hybridization of O 2p with the highly dispersed Zn 4s states, in which the original Peak 3 (~ 237.8 eV) can actually be considered as the bottom of the conduction band.[8, 9] On this basis, Peak 1 may not fundamentally be regarded as a single and separate peak, since in this case it appears that all the 3d orbitals of Zn are full, and the spectrum should show only the O 2p states that hybridises with the 4sp states of Zn.[10]

Analogously, the Peak 4 (~ 541.1 eV) and 5 (~ 543.8 eV) in the energy range between 539 and 550 eV should be assigned to the hybridization of the O 2p states with the Zn 4p states.[8, 9] Regarding the potential Peak 6 located around 556.4 eV, it may be caused by the extension of the O 2p states to higher orbitals of Zn, such as the Zn 4d states.[8, 9]

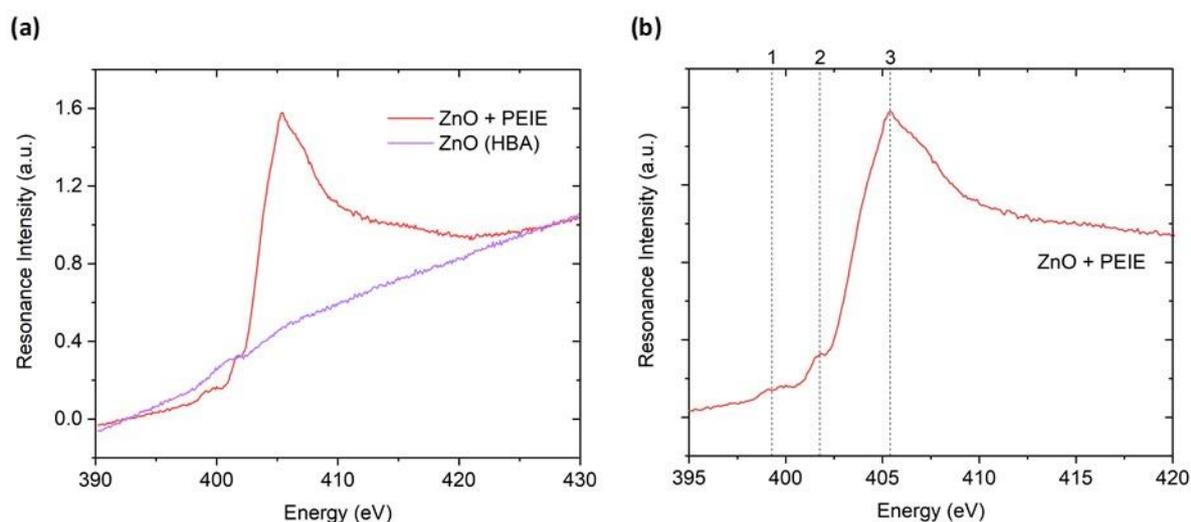

*Figure S8:* N 1s K-edge NEXAFS spectra with peak identification for each spin-coated film sample at 55° X-ray incidence. In part *(a)* all spectra are overlaid, while in part *(b)* the spectra for each sample are offset for clarity.

| Peak No. | Energy Position (eV) | Peak Notation |
|---|---|---|
| 1 | ~ 399.1 | N.A. |
| 2 | ~ 401.9 | N 1s → σ* (N–H) |
| 3 | ~ 405.3 | N 1s → σ* (N–C) |

*Table S8:* Peak assignments and energy positions for N 1s K-edge NEXAFS spectra at 55° X-ray incidence.

Figure S8 and Table S8 show the NEXAFS spectra versus peak assignments of the K-edge for N 1s at an angle of incidence of 55° X-rays, respectively. It is worth noting that only the spectra of the sample with PEIE applied are analysed here. Although the HBA modified sample theoretically also contains nitrogen (see Figure 1), the amount is so small that it does not produce sufficient signal to form a discernible peak (see Figure S8 (a)). Intuitively, there is only one obvious peak in these spectra, whereas, if the distinct fluctuations are also considered, the three constituent peaks listed in the Table S8 can be roughly obtained. Although there is a relatively distinctive raised feature at the location of Peak 1 (~ 399.1 eV) for the PEIE coated sample, its corresponding signal origin cannot be found at present. Considering that the nitrogen K-edge is supposed to occur at around 400 eV, such feature may be just an arbitrary measurement fluctuation.[7] For the Peak 2 and Peak 3 with energy positions of approximately 401.9 eV and 405.3 eV, respectively, their resonance energies should both come from the N 1s → σ* transition, except that the former should be related to N–H bonds, while the latter should be related to N–C bonds.[11, 12]

A schematic diagram of an angle-resolved NEXAFS experiment is shown in Figure S9, for the sake of discussion, only a single benzene ring without side chains is considered herein as the object being probed. A beam of X-rays is incident in the *x-z* plane at an angle of incidence ($\theta$), and then the polarisation vector of the beam (*E*) also appears in this plane provided that only perfectly linearly polarised X-rays (as generated by undulator beamlines) are considered. If the *x-y* plane is defined as the substrate plane, the tilt angle ($\alpha$) of the benzene ring can then be expressed as the angle between the surface normal along the *z* axis, and the transition dipole moment (*O*) orientated perpendicular to the molecular plane. Thus, a value of $\alpha$ equal to 90° means that the benzene ring is completely "edge-on" with respect to the substrate, while a value of α equal to 0° means that the benzene ring is completely "face-on" with respect to the substrate.[7]

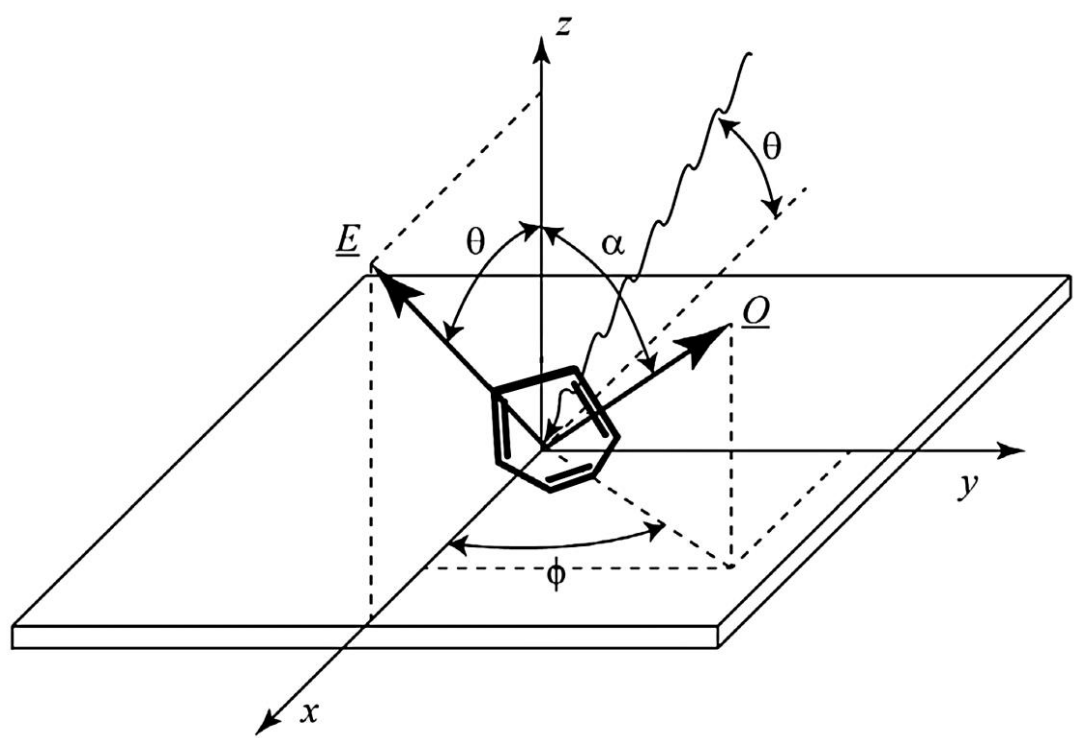

*Figure S9: The geometry of an angle-resolved NEXAFS spectroscopy experiment for determining the molecular orientation of an aromatic (benzene) ring with respect to the plane of the substrate, where x-y plane is the substrate plane, z axis is the surface normal, E is the polarisation vector of the beam, O is the transition dipole moment, $\vartheta$ is the angle of incident, $\varphi$ is the azimuthal angle, and α is the tilt angle.[7] Adapted with permission from M. M. Nahid, E. Gann, L. Thomsen, and C. R. McNeill, in European Polymer Journal, vol. 81 (Elsevier, 2016), p. 538. Copyright 2016 Elsevier. Published by Elsevier Ltd.*

For spin-coated films such as those used in this experiment, averagely, the chains of molecules or polymers are not preferentially aligned in any direction along the substrate plane. Besides, the size of the X-ray beams of NEXAFS spectroscopy is generally larger than the single crystal domains, which means that the beams are practically probing an ensemble of the chains as well as sampling the backbone conformations with large distributions. In other words, the beams theoretically sample the benzene ring with every possible value of the azimuthal angle ($\varphi$), thereby averaging out the $\varphi$-dependence of $\underline{E}\cdot\underline{O}$.[7] In the case of the sample is rotationally symmetric around the substrate normal, the intensity of the resonance (*I*) excited is decided by the angle between the electric field vector of the incident beam ($\underline{E}$) and the transition dipole moment ($\underline{O}$), that is, the maximum *I* is only obtained when the $\underline{O}$ is parallel to the $\underline{E}$.[13] Therefore, the orientation of the molecular plane with respect to the surface plane of the sample can be confirmed by the variation of the resonance intensity (*I*) of the electron excitation from carbon 1s orbital to the aromatic $\pi$ * level (i.e. the C 1s → $\pi$* transition), with the incident angle ($\theta$) of photons.[7, 13] For a given $\theta$, this $\pi$* resonance intensity (*I*) is solely dependent on the tilt angle ($\alpha$) of the aromatic conjugated backbone relative to the surface normal. Such $\theta$-dependence of *I* is known as dichroism, and hence it is only needed to measure $I(\theta)$ for any arbitrary azimuthal angle ($\varphi$) to describe the tilt angle ($\alpha$) of the system, which can be expressed as the following equation[7, 13]:

$$I(\theta) = TRI \frac{1}{3}\left[1 + \frac{1}{2}(3\cos^2\theta - 1)(3\cos^2\alpha - 1)\right] \qquad \text{Equation 1}$$

Where *TRI* is the total resonance intensity, namely the intensity measured for $\underline{E}\cdot\underline{O} = 1$.[7]

The angle-resolved NEXAFS spectra of the BZA derivatives modified ZnO spin coated films are shown in the images (a), (c), (e) of Figure S10, which were all recorded at X-ray incidence angles ($\theta$) of 20°, 40°, 55°, 70° and 90°. It is worth bearing in mind that these energy-dependent partial X-ray absorption cross-section spectra are actually normalised, in which the intensity values at the pre-edge (~ 280 eV) and post-edge (~ 320 eV) of the spectra are set to around 0 and 1 respectively.[14] The pre-edge is conventionally set obviously before the onset of the first resonance transition in the spectrum; while the post-edge should be determined well above the ionisation threshold (i.e. at the continuum), where there is no longer any resonant transition and associated dichroism.[7, 14] In addition, it may be observed that the pre-edge background of the spectra in Figure S10 is somewhat decaying, which is assumed to be essentially flat in theory. This may be related to the absorptions from lower energy transitions in the practical testing.[7]

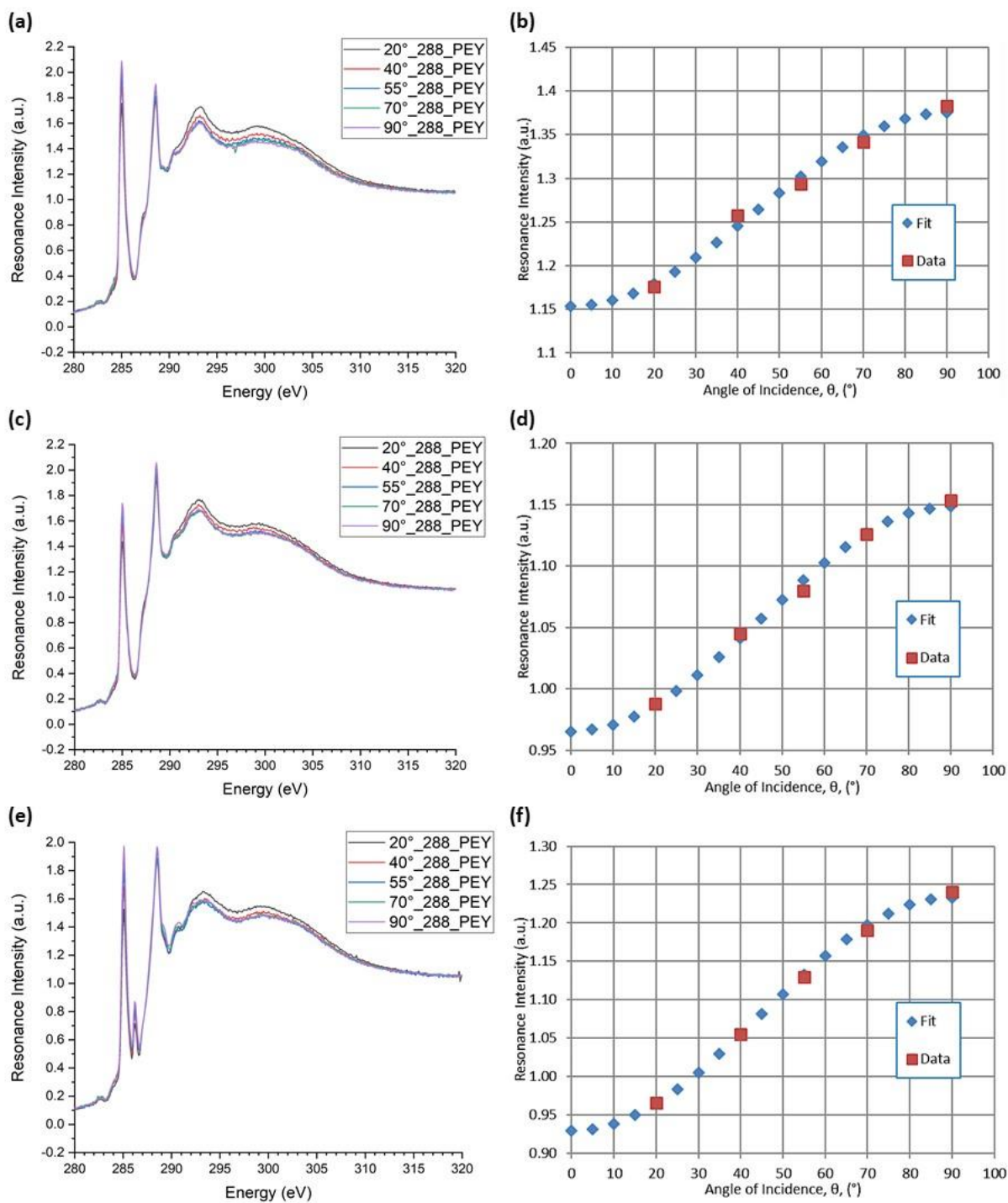

*Figure S10:* Angle-resolved NEXAFS spectra of BZA *(a)*, HBA *(c)*, CBA *(e)* modified film samples with pre-edge (~ 280 eV) set to 0 and post-edge (~ 320 eV) set to 1, and their corresponding dichroism plots *((b), (d), (f))* of the C 1s → π* transition resonance intensity at ~ 285 eV as a function of ϑ.

The images (b), (d), (f) in the Figure S10 presents the correlated dichroism plots of the angle-resolved NEXAFS spectra (a), (c), (e), respectively, which focus specifically on the C 1s → π* transition at about 285 eV, as this transition is directly related to the aromatic ring being probed (see Table 3). By fitting the sets of data to the Equation 1, it can be determined that the tilt angles ($α$) for the BZA, HBA, and CBA modified samples are around 57.1°, 57.0° and 58.5°, respectively. As mentioned earlier, the values of $α$ obtained here essentially represent the ensemble-averaged tilt angles. Considering that these spin coated films potentially possess a wide conformation range as well as a large tilt angle distribution, if the resulting $α$ value is an intermediate value that is neither close to 90° for the "edge-on" nor close to 0° for the "face-on" configuration, it is actually ambiguous that whether the probed molecular planes (i.e. the benzene rings) are predominantly oriented with the intermediate tilt angle, or whether there is a wide disordered distribution of tilt angles but without a high proportion of "edge-on" and "face-on" configurations.[7] Nevertheless, for the case of rotationally symmetric spin-coated films herein, there still exists a viable empirical criterion that the tilt angle ($α$) coincides with zero dichroism is approximately 54.7°, which is sometimes referred to as the "magic angle".[7] Since the measured $α$ results for all samples here are greater than this threshold value and relatively close to 90°, the orientation of the molecular backbone (i.e. benzene ring) of these BZA derivatives on the surface of the ZnO substrate is statistically more inclined to be "edge-on", although not significantly so.

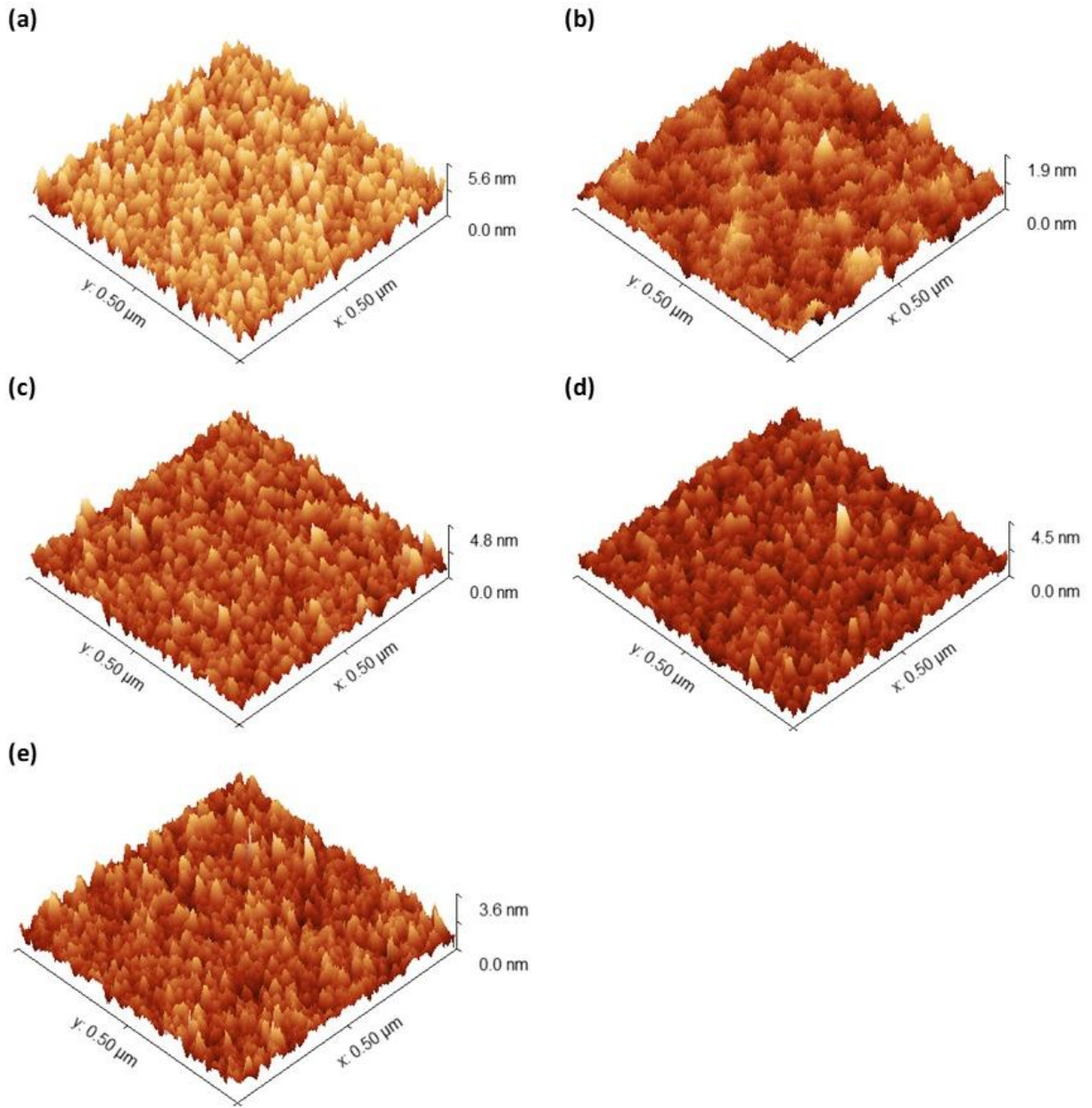

***Figure S11:*** *Thin film surface three-dimensional AFM images of **(a)** ZnO (Reference); **(b)** ZnO (Reference) + PEIE; **(c)** ZnO (BZA Modified); **(d)** ZnO (CBA Modified); **(e)** ZnO (HBA Modified).*

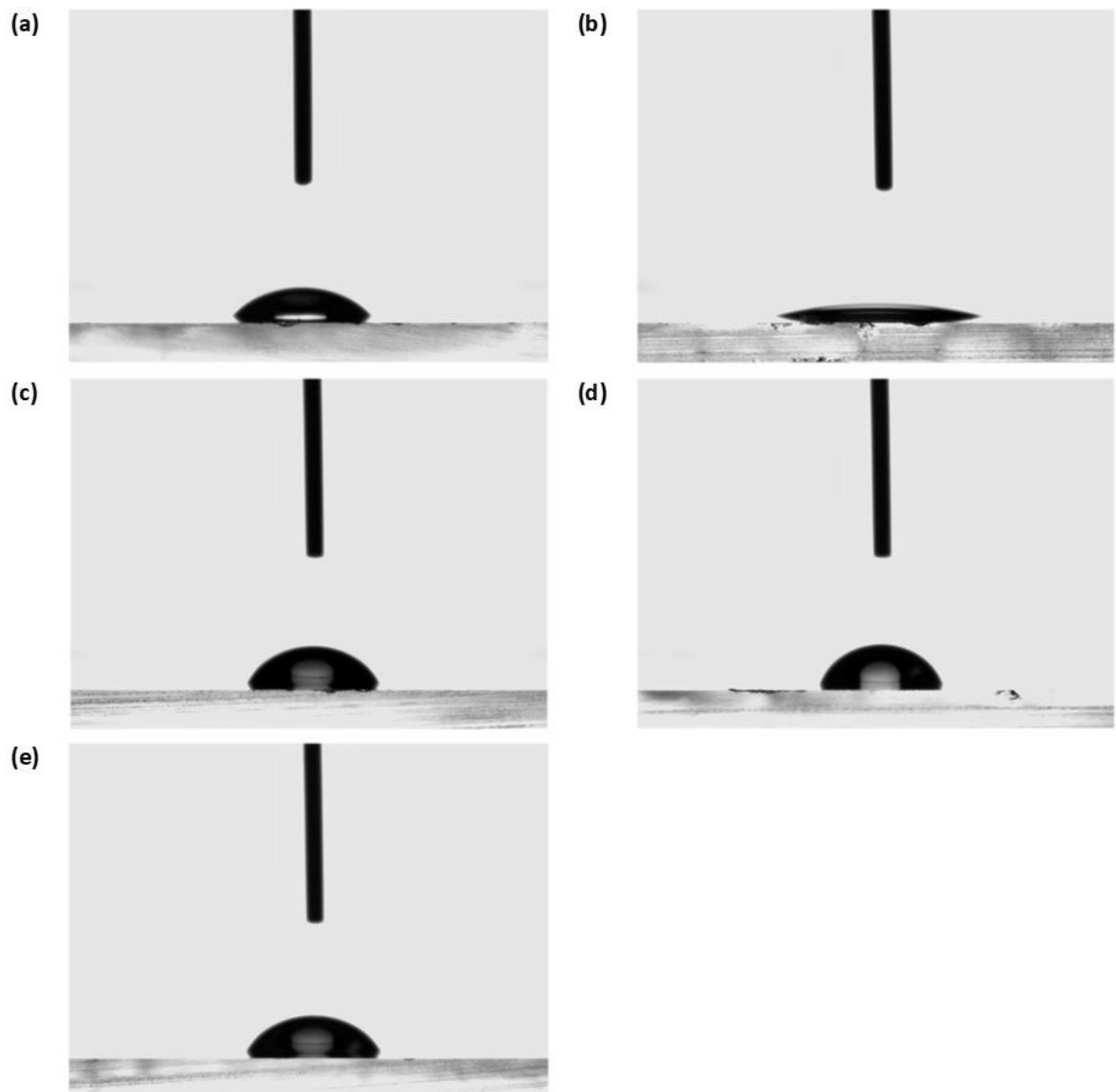

*Figure S12: Representative digital photographs from water contact angle measurements on **(a)** ZnO (Reference); **(b)** PEIE; **(c)** ZnO (BZA Modified); **(d)** ZnO (CBA Modified); **(e)** ZnO (HBA Modified).*

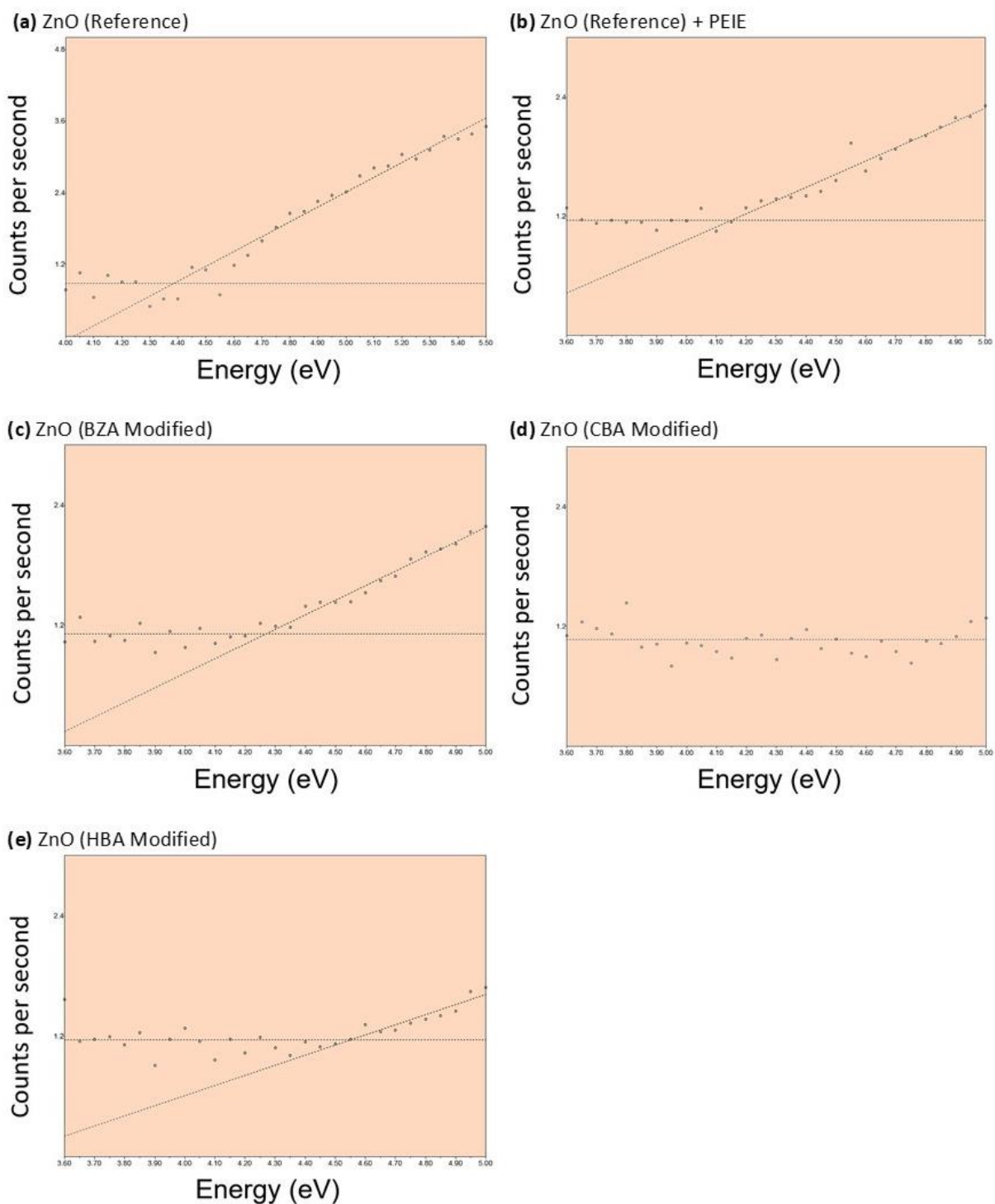

***Figure S13:*** *Scatter plots of PESA data examples have a vertical axis representing the square root of the electron yield and a horizontal axis representing the photon energy. The energy threshold for photoemission, which corresponds to the ionization potential and ultimately the work function, is determined by the point at which the background fitting line intersects with the extrapolated fitting line of the linear segment of the square root plots of photoemission yield.*

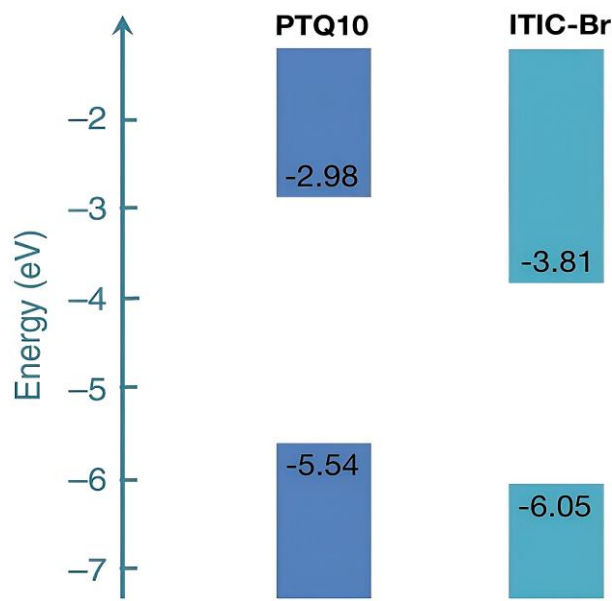

***Figure S14:*** *Energy level diagram of the active layer materials used in the OSCs, where values are with respect to the vacuum level and negative values indicate that energy has to be added to remove the bound electrons. Data for ITIC-Br measured using cyclic voltammetry by Mr. Doan Van Vu, and data for PTQ10 from the reference.*[15]

These WF results can also be considered in light of the photovoltaic measurements. For instance, the photovoltaic performance of the PEIE modified and BZA modified devices are quite similar, see Table 1, with the same $V_{OC}$ value (0.92 V), which is greater than that of the reference ZnO device (0.88 V). The $V_{OC}$ dependence is expected to be correlated with the WF of the ETL. As shown in Figure S14, the HOMO and LUMO of the PTQ10 donor are known to be located at approximately 5.54 eV and 2.98 eV, respectively. The preferential energy level alignment in the architecture of OSCs suggests that the WF level of the ZnO ETL should be as close as possible to the LUMO level of PTQ10 and far away from its HOMO level, to ensure efficient electron flow and minimise hole recombination at the cathode.[16, 17] Thus, a smaller WF value of the ETL would bring the quasi-Fermi level of the electrons ($E_{F,e^-}$) of ZnO closer to the LUMO of PTQ10, resulting in a larger built-in potential ($V_{bi}$), which creates the internal electric field to generate photocarriers, and hence a higher $V_{OC}$ can be obtained.[18-21]

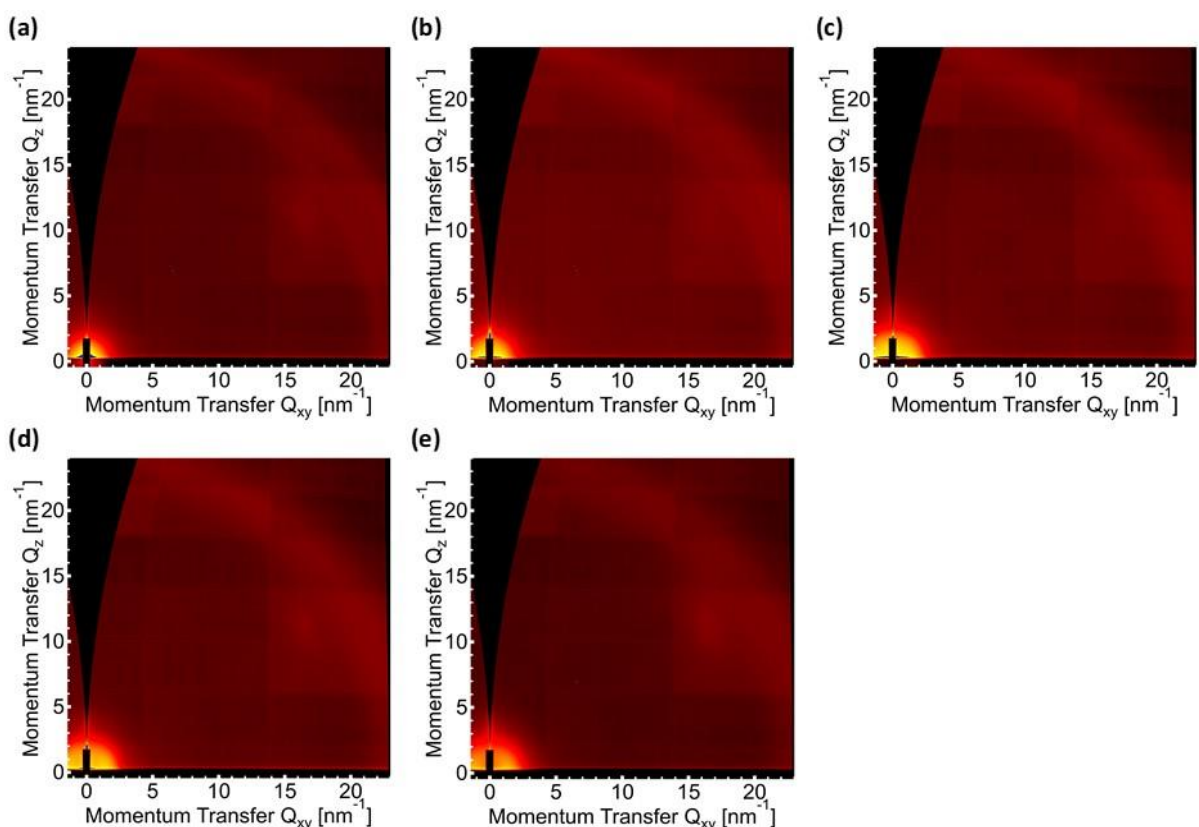

*Figure S15:* Two-dimensional GIWAXS patterns of *(a)* Si / ZnO (Ref.); *(b)* Si / ZnO (Ref.) / PEIE; *(c)* Si / ZnO (BZA Mod.); *(d)* Si / ZnO (CBA Mod.) and *(e)* Si / ZnO (HBA Mod.). The colour coding visually shows the difference in scattering intensity. All GIWAXS patterns displayed here have been corrected for geometrical distortion of reciprocal space captured in raw patterns resulting in the missing wedge along $Q_{xy} = 0$ nm$^{-1}$.

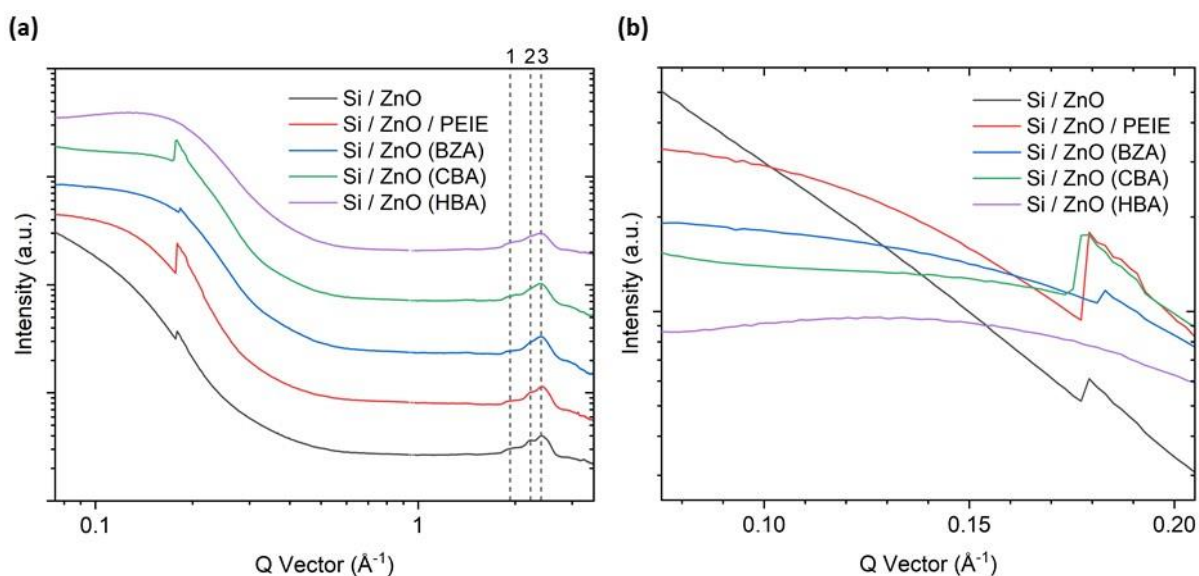

*Figure S16: (a)* Reduced one dimensional scattering profiles obtained by radial averaging of the 2D GIWAXS patterns, the profiles are offset for clarity; *(b)* Low scattering vector region of reduced, non-offset one dimensional scattering profiles. The discontinuity at ~ 0.2 Å$^{-1}$ is due to the beam stop.

It is well known that small-angle X-ray scattering (SAXS) data can provide some insights into the conformation of particles or aggregates at the nanoscale, including the radius of gyration of scatterers or called Guinier radius ($R_g$), the Porod volume ($V_p$), the correlation volume or length, and the surface-to-volume ratio, etc.[22, 23] However, quantitative analysis of the SAXS region is limited by the $Q$ range probed (does not extend to low enough $Q$), which prevents further quantitative analysis. Proper analysis of GISAXS data also requires the treatment of multiple scattering effects which is beyond the scope of this article. Nevertheless, one can see that the ZnO reference sample shows enhanced scattering at lower $Q$ relative to the other samples, see Figure S16 (b), which suggests a larger particle size. In the other samples, shoulders or even peaks can be seen between 0.1 and 0.2 Å$^{-1}$ indicative of smaller sized particles than found in the ZnO reference. More accurate determination of particle size requires higher quality data of dedicated GISAXS studies for studying such samples.

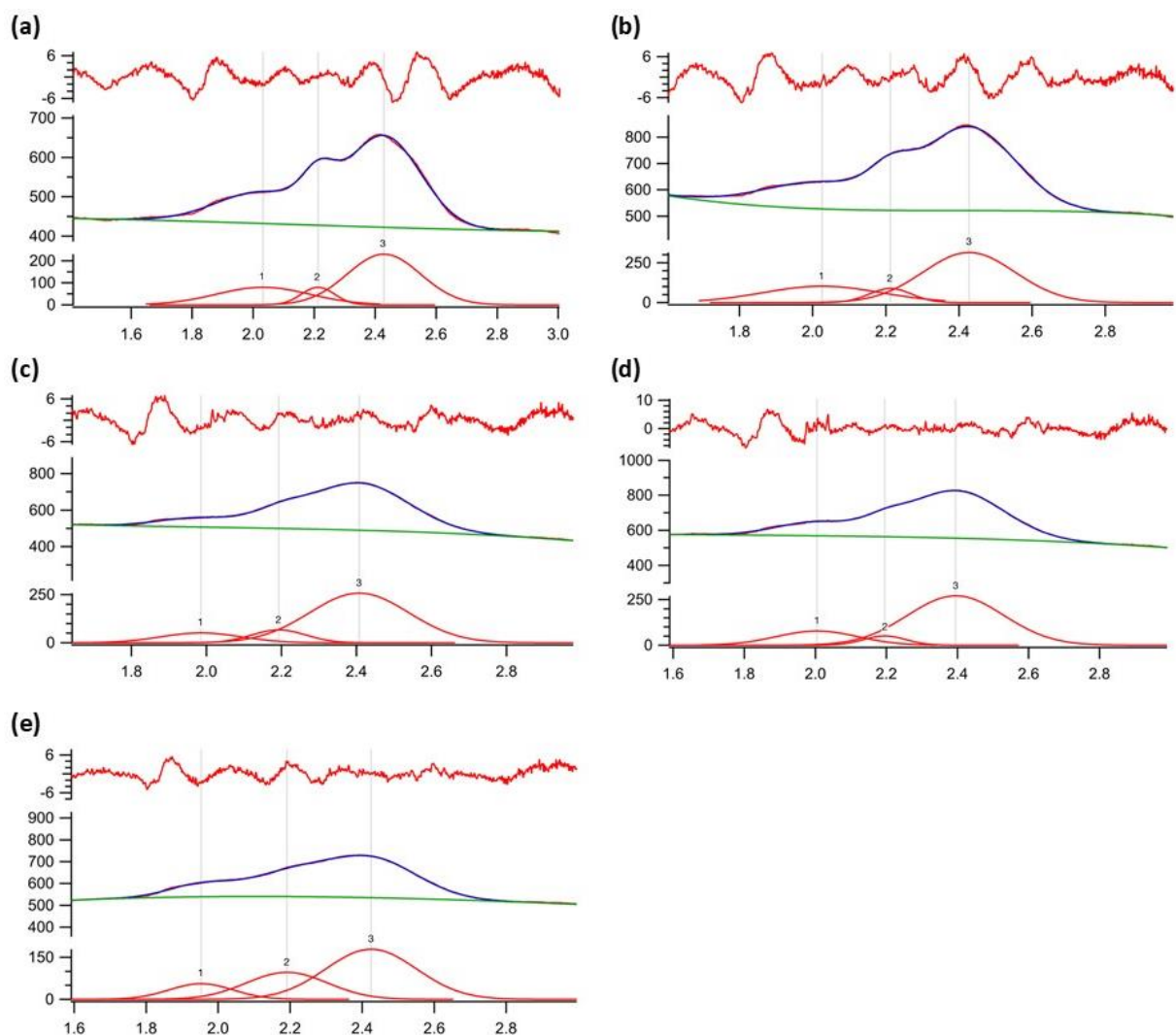

*Figure S17:* Fittings of 1D scattering profiles in high Q region for *(a)* Si / ZnO (Ref.); *(b)* Si / ZnO (Ref.) / PEIE; *(c)* Si / ZnO (BZA Mod.); *(d)* Si / ZnO (CBA Mod.); *(e)* Si / ZnO (HBA Mod.).

| Miller index (h k l) | Standard 2θ (Deg.) | Q Vector (Å$^{-1}$) |
|---|---|---|
| 1 1 1 | 28.443 | 2.005 |
| 2 2 0 | 47.303 | 3.274 |
| 3 1 1 | 56.123 | 3.839 |
| 4 0 0 | 69.131 | 4.630 |
| 3 3 1 | 76.377 | 5.045 |
| 4 2 2 | 88.032 | 5.670 |
| 3 3 3 | 94.954 | 6.014 |
| 4 4 0 | 106.710 | 6.547 |
| 5 3 1 | 114.094 | 6.847 |
| 6 2 0 | 127.547 | 7.320 |
| 5 3 3 | 136.897 | 7.590 |
| 4 4 4 | 158.638 | 8.019 |

**Table S9:** *Position of diffraction peak positions of Si powder, wavelength of the X-ray radiation is 1.54 Å.*[24]

| Miller index (h k l) | Standard 2θ (Deg.) | Q Vector (Å$^{-1}$) |
|---|---|---|
| 1 0 0 | 31.769 | 2.233 |
| 0 0 2 | 34.421 | 2.414 |
| 1 0 1 | 36.252 | 2.539 |
| 1 0 2 | 47.538 | 3.289 |
| 1 1 0 | 56.602 | 3.869 |
| 1 0 3 | 62.862 | 4.255 |
| 2 0 0 | 66.378 | 4.467 |
| 1 1 2 | 67.961 | 4.561 |
| 2 0 1 | 69.100 | 4.628 |

**Table S10:** *Position of diffraction peak for ZnO powder, wavelength of the X-ray radiation is 1.54 Å.*[25, 26]

| Sample | FWHM (Å$^{-1}$) | |
|---|---|---|
| | Peak 2 (ZnO (100)) | Peak 3 (ZnO (002)) |
| Si / ZnO (Ref.) | 0.13 (± 0.01) | 0.28 (± 0.01) |
| Si / ZnO (Ref.) / PEIE | 0.14 (± 0.01) | 0.30 (± 0.01) |
| Si / ZnO (BZA Mod.) | 0.17 (± 0.01) | 0.31 (± 0.01) |
| Si / ZnO (CBA Mod.) | 0.14 (± 0.01) | 0.32 (± 0.01) |
| Si / ZnO (HBA Mod.) | 0.27 (± 0.01) | 0.30 (± 0.01) |

**Table S11:** *Summary of fitted FWHM for ZnO (100) and (002) peaks as probed using GIWAXS.*

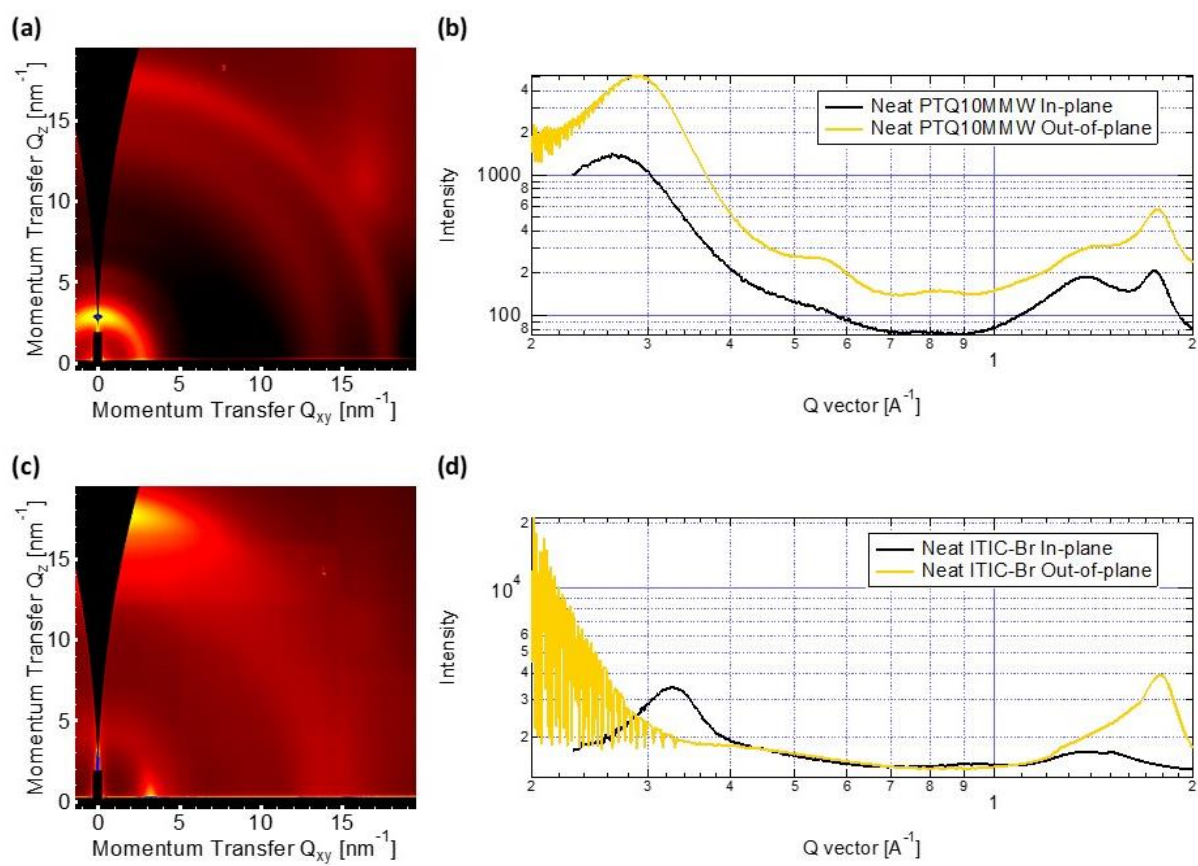

*Figure S18:* Two-dimensional GIWAXS patterns of neat **(a)** PTQ10 and **(c)** ITIC-Br, the corresponding reduced one-dimensional scattering profiles are plotted in **(b)** and **(d)** respectively.

Before discussing PTQ10:ITIC-Br films prepared on different ETL samples, it is necessary to have an understanding of the crystalline properties of the active layer itself first. In the neat PTQ10 film, as shown in Figure S18 (a) & (b), the (100) lamellar stacking and (010) π-π stacking peaks are located at 0.29 Å$^{-1}$ (2.9 nm$^{-1}$) and 1.77 Å$^{-1}$ (17.7 nm$^{-1}$), respectively. Although there seem to be no obvious primary orientation direction for these two kinds of stacking in PTQ10 as they still show "ring-like" features in the 2D GIWAXS patterns, both lamellar stacking and π-π stacking peaks are relatively strong in the out-of-plane direction and weak in the in-plane direction. In particular, the stronger out-of-plane π-π stacking diffraction suggests that the molecular stacking in the vertical direction of the substrate tends to be more face-on oriented, which is favourable for charge transport.[15] Moreover, a "wide halo" can be observed at approximately 1.4 Å$^{-1}$ (14 nm$^{-1}$) in between the well-defined (100) and (010) peaks and can also be seen on the corresponding 1D profiles. This broad feature is attributed to amorphous PTQ10 chains, and is seen in many carbon-based materials, and not necessarily only in the PTQ10 studied here.[27]

Regarding neat ITIC-Br (see Figure S18 (c), (d)), the film exhibit (100) lamellar stacking and (010) π-π stacking peaks at 0.33 Å$^{-1}$ (3.3 nm$^{-1}$) and 1.79 Å$^{-1}$ (17.9 nm$^{-1}$), respectively. Notably, these two peaks appears textured where the lamellar peak is stronger along the in-plane direction, and the π-π stacking peak is stronger along the out-of-plane direction as depicted in the reduced 1D plots. As such, it can be inferred that the ITIC-Br crystallites in the film adopt a predominant face-on orientation with respect to the substrate. This observation is also good evidence of the potential of this small molecule in OPV applications, as it has been demonstrated that face-on crystalline orientation can better facilitate efficient charge transport in OSCs.[28-30]

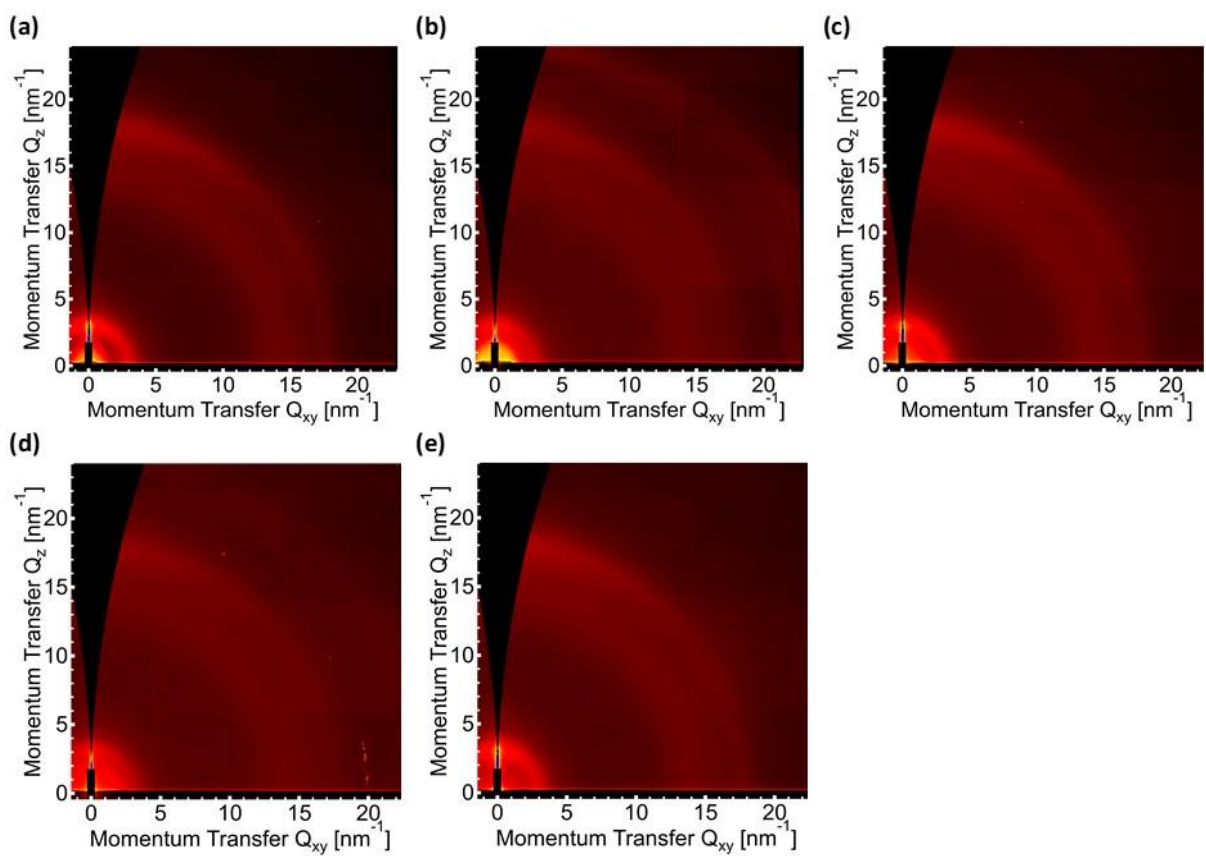

*Figure S19:* Two-dimensional GIWAXS patterns of *(a)* Si / ZnO (Ref.) / PTQ10:ITIC-Br; *(b)* Si / ZnO (Ref.) / PEIE / PTQ10:ITIC-Br; *(c)* Si / ZnO (BZA Mod.) / PTQ10:ITIC-Br; *(d)* Si / ZnO (CBA Mod.) / PTQ10:ITIC-Br, and *(e)* Si / ZnO (HBA Mod.) / PTQ10:ITIC-Br. The colour scale visually shows the difference in scattering intensity.

It is concluded based on the GIWAXS results for the neat active layer materials that crystallites of both the electron donor (PTQ10) and acceptor (ITIC-Br) prefer a face-on orientation in their own neat thin film, especially the latter. Interestingly, it can be visualised that the microstructural features of these individual components are largely retained in the blend films

S29

of PTQ10 and ITIC-Br, regardless of the ETL being investigated. As shown in the 2D GIWAXS patterns in Figure S19, a relatively stronger preference for face-on orientation can still be observed in all samples, which could be one of the reasons that PTQ10:ITIC-Br system affords a good photovoltaic performance, just as previously described. Nevertheless, it can also be seen that the structure within the PTQ10:ITIC-Br blend films becomes less ordered compared to the component neat films, as the crystalline features are considerably suppressed, especially the in-plane features, although this also in a sense indicates the relatively good miscibility of PTQ10 with ITIC-Br.

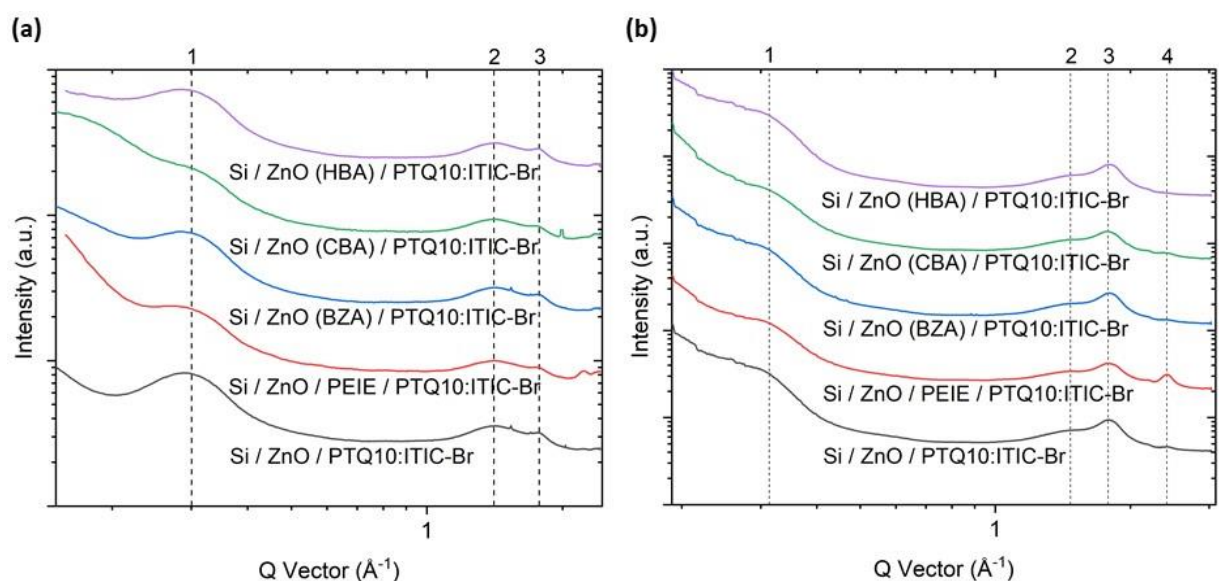

*Figure S20:* Offset reduced one dimensional scattering profiles in logarithmic scales obtained by azimuthal integration along **(a)** in-plane direction (ca. 0 ~ 5°) and **(b)** out-of-plane direction (ca. 85 ~ 95°) of radial slices from the 2D GIWAXS patterns.

The crystallographic properties of these blend samples can be better interpreted via the corresponding reduced 1D scattering profiles for the various samples, see Figure S20. Note that unlike the previously discussed bare ETL samples, there is a discernible difference between features in the in-plane azimuthal range and out-of-plane azimuthal range of the samples covered by active layers (see Figure S19), thus, it is necessary to analyse the 1D profiles of the in-plane average and out-of-plane average for each sample separately. The summarised in-plane and out-of-plane scattering profiles are shown in Figure S20 (a) and (b), with the scattering profiles for each sample presented in Figure S21.

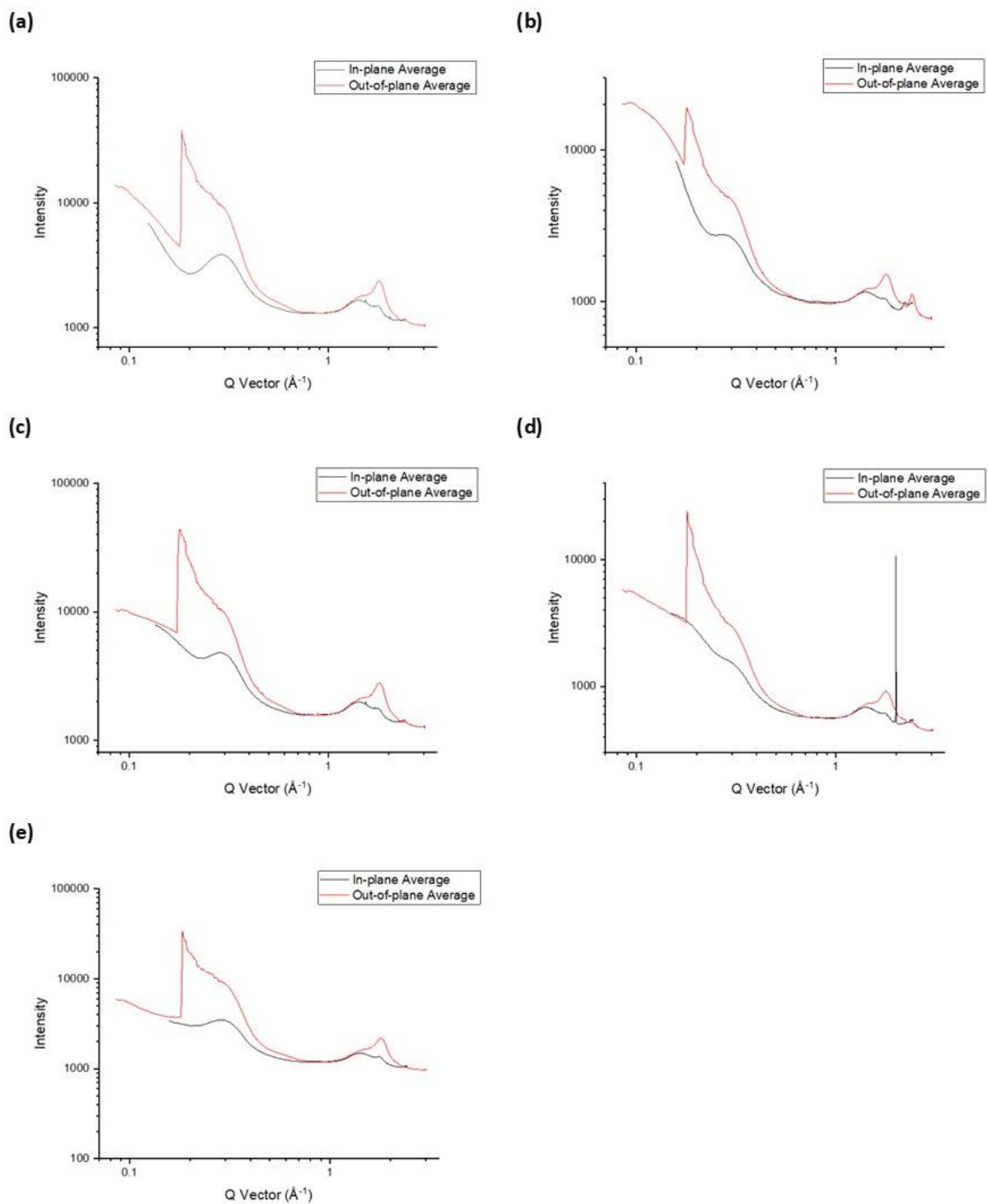

***Figure S21:*** *Reduced 1D scattering profiles in logarithmic scales of **(a)** Si / ZnO (Ref.) / PTQ10:ITIC-Br; **(b)** Si / ZnO (Ref.) / PEIE / PTQ10:ITIC-Br; **(c)** Si / ZnO (BZA Mod.) / PTQ10:ITIC-Br; **(d)** Si / ZnO (CBA Mod.) / PTQ10:ITIC-Br, and **(e)** Si / ZnO (HBA Mod.) / PTQ10:ITIC-Br.*

| Sample | Q Vector (Å⁻¹) | | |
|---|---|---|---|
| | Peak 1 | Peak 2 | Peak 3 |
| Si / ZnO (Ref.) / PTQ10:ITIC-Br | 0.30 (± 0.01) | 1.43 (± 0.01) | 1.78 (± 0.01) |
| Si / ZnO (Ref.) / PEIE / PTQ10:ITIC-Br | 0.30 (± 0.01) | 1.42 (± 0.01) | 1.77 (± 0.01) |
| Si / ZnO (BZA Mod.) / PTQ10:ITIC-Br | 0.30 (± 0.01) | 1.43 (± 0.01) | 1.78 (± 0.01) |
| Si / ZnO (CBA Mod.) / PTQ10:ITIC-Br | 0.30 (± 0.01) | 1.43 (± 0.01) | 1.77 (± 0.01) |
| Si / ZnO (HBA Mod.) / PTQ10:ITIC-Br | 0.30 (± 0.01) | 1.43 (± 0.01) | 1.77 (± 0.01) |
| Peak Assignment | PTQ10 (100) ITIC-Br (100) | Amorphous Carbon | PTQ10 (010) |

***Table S12:*** *Summary of diffraction peak locations and assignments for GIWAXS measurements of various ETLs with PTQ10:ITIC-Br on top (in-plane average).*

| Sample | Q Vector (Å⁻¹) | | | |
|---|---|---|---|---|
| | Peak 1 | Peak 2 | Peak 3 | Peak 4 |
| Si / ZnO (Ref.) / PTQ10:ITIC-Br | 0.30 (± 0.01) | 1.48 (± 0.01) | 1.80 (± 0.01) | 2.44 (± 0.02) |
| Si / ZnO (Ref.) / PEIE / PTQ10:ITIC-Br | 0.31 (± 0.01) | 1.47 (± 0.01) | 1.80 (± 0.01) | 2.42 (± 0.01) |
| Si / ZnO (BZA Mod.) / PTQ10:ITIC-Br | 0.31 (± 0.01) | 1.53 (± 0.01) | 1.81 (± 0.01) | - |
| Si / ZnO (CBA Mod.) / PTQ10:ITIC-Br | 0.31 (± 0.01) | 1.50 (± 0.01) | 1.79 (± 0.01) | - |
| Si / ZnO (HBA Mod.) / PTQ10:ITIC-Br | 0.31 (± 0.01) | 1.55 (± 0.01) | 1.81 (± 0.01) | - |
| Peak Assignment | PTQ10 (100) ITIC-Br (100) | Amorphous Carbon | ITIC-Br (010) | Uncertain |

***Table S13:*** *Summary of diffraction peak locations and assignments for GIWAXS measurements of various ETLs with PTQ10:ITIC-Br on top (out-of-plane average).*

According to Figure S20, Table S12 and Table S13, in general, there are not significant differences in the microstructural features of the PTQ10:ITIC-Br active layer prepared on the different substrates. Quantitative analysis of the (100) peak is hampered by the overlapping nature of the (100) peak from PTQ10 and ITIC-Br. The (100) peak of ITIC-Br is not clearly resolved, which indicates that ITIC-Br is less order in blends with PTQ10. The peak positions determined (see Table S12 and Table S13) do not vary appreciably, and differences seen at low $Q$ about the (100) peak (i.e. Peak 1) can be attributed to variations in SAXS due to roughness and not due to differences in crystalline properties of the blend. There is also not much difference seen in the texture of the different samples (preferential face-on vs. edge-on orientation). In addition, an additional peak (Peak 4) is seen at high $Q$ for PTQ10:ITIC-Br active layers on ZnO (Ref.) and ZnO (Ref.) / PEIE; the origin of this peak is still uncertain as it does not appear in neat films of PTQ10 or ITIC-Br nor in the bare ETL samples.

It is generally acknowledged that in conjugated materials, the charge transport depends primarily on π-orbital overlap, because the conjugated structure allows electrons to delocalise in all adjacent aligned π-orbital, and then spread over a larger area, bringing added stability to the whole structure.[31] For conjugated polymers, of the three major crystallographic axes - backbone, π-π stacking and lamellar stacking - the highest charge transport mobility is possessed along the first two directions.[7, 32] When considering conjugated polymers for OSCs, a "face-on" molecular arrangement is usually preferred rather than a "edge-on" one, as the efficient charge transport π-π stacking direction is the same as the out-of-plane charge transport direction in OSCs.[31, 33] Consequently, the good performance of the device based on PTQ10 and ITIC-Br active layer system can be attributed to the fact that the crystallites of both materials prefer a "face-on" orientation in the blend film, resulting from the stronger π-π stacking along the out-of-plane direction.

## ITIC-Br synthesis

IT-CHO (374 mg) and IC-Br (206 mg) were dissolved in a mixture of chloroform with 2% pyridine (100 mL). The resulting mixture was refluxed under nitrogen for 22 hours. The solvent was removed in vacuo and the crude was purified by column chromatography with chloroform, yielding the product as a dark blue powder (501 mg, 90.9%).

## IC-Br

5-bromo-1,3-indandione (4.00 g) and malononitrile (3.60 g) were added to a round bottom flask under nitrogen. Ethanol (30 mL) and anhydrous sodium acetate (1.94 g) was added to the mixture, which immediately turned red, and the resulting solution was stirred at room temperature for 40 minutes. The reaction mixture was poured into water (150 mL) and the pH was adjusted to 1-2 by the addition of hydrochloric acid. The formed solid was filtered off, dissolved in acetone, and dry loaded on celite. The dry loaded crude was then further purified by column chromatography (shielded from light with Al-foil) with a solvent gradient of DCM/hexane 9:1 to pure DCM, yielding the product as a powder (2.45 g, 29.8%).

$^1$H-NMR spectra were measured in CDCl$_3$ using a Bruker 600 MHz NMR spectrometer. The recorded $^1$H-NMR spectra were referenced to the residual CHCl$_3$ solvent peaks.

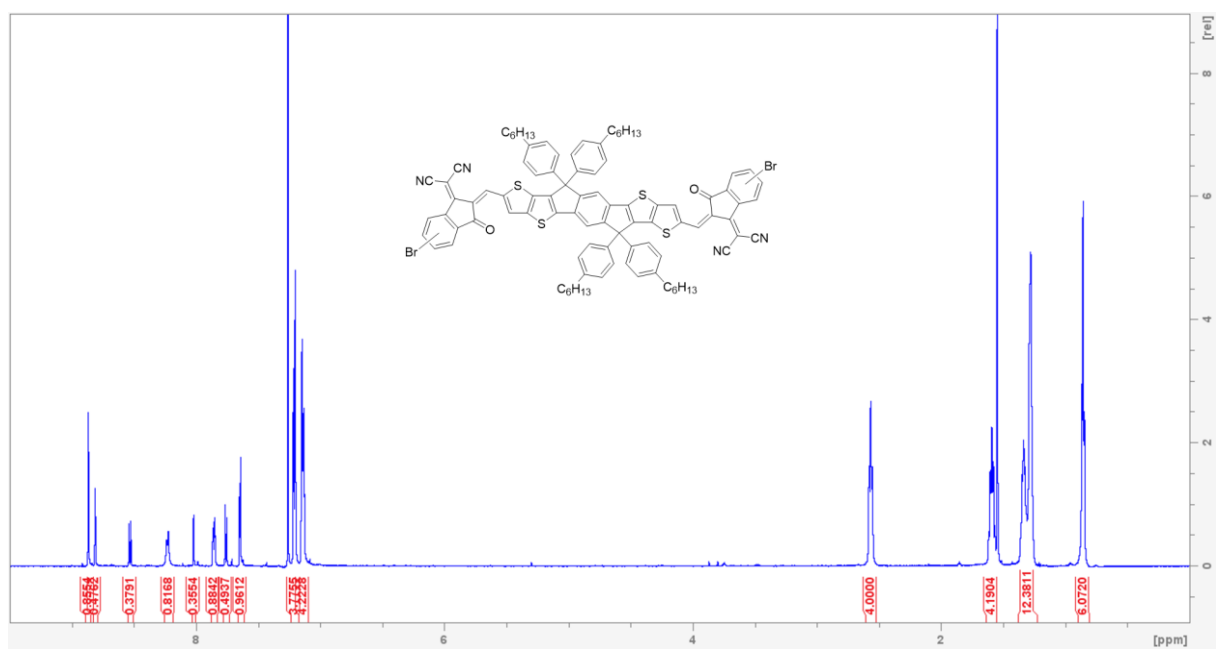

*Figure 22.* NMR of ITIC-Br.